\documentstyle[12pt]{article}
\begin{document}
\begin{center}
{\bf Coding Theorems for Quantum Channels}
\vskip20pt

A. S. Holevo\\ Steklov Mathematical Institute
\end{center}\vskip20pt

{\bf\small Abstract -- The more than thirty years old issue of the
(classical) information capacity of quantum communication channels was
dramatically clarified during the last years, when a number of direct quantum
coding theorems was discovered. The present paper gives a self contained
treatment of the subject, following as much in parallel as possible with
classical information theory and, on the other side, stressing profound
differences of the quantum case. An emphasis is made on recent results, such
as general quantum coding theorems including cases of infinite (possibly
continuous) alphabets and constrained inputs,
reliability function for pure state channels
 and quantum Gaussian channel. Several still unsolved problems
are briefly outlined.}
\vskip20pt
\centerline{\sc I. Introduction}
\vskip10pt
The issue of the information capacity of quantum communication channels arose
in the sixties (see, in particular, \cite{xgordon62},\cite{xforney},
\cite{xle}, \cite{xgordon}  and more references in the survey \cite{xcaves}) and goes back
to even earlier classical papers of Gabor and Brillouin, asking for
fundamental physical limits on the rate and quality of information
transmission. This work laid a physical foundation and raised the question of
consistent quantum information treatment of the problem. Important steps in
this direction were made in the seventies when quantum statistical decision
(detection and estimation) theory \cite{xhel}, \cite{xstat} was created,
making a quantum probabilistic frame for this circle of problems. At that
time the quantum entropy bound and strict superadditivity of classical 
information in
quantum communication channels were established \cite{xhol73}, \cite{xhol79}.

A substantial progress has been achieved during the past two years, when a number
of direct quantum coding theorems was discovered, proving the achievability
of the entropy bound \cite{xjozsa}, \cite{xhol}, \cite{xwest}. To
considerable extent this was stimulated by an interplay between the quantum
communication theory and quantum information ideas related to more recent
development in quantum computing (see e.g.  \cite{xben}). The question of
information capacity is important in the theory of quantum computer, which is
a highly specific information processing device, particularly in connection
with quantum error-correcting codes
\cite{xcal}, \cite{xste}.

In this paper we discuss transmission of classical information through
quantum channels. Remarkably, important probabilistic tools underlying the
treatment of this case have their roots, and in some cases direct
prototypes, in classical Shannon's theory, as presented in particular in
\cite{xgal}, \cite{xinf}.
The paper is intended to give a self contained and rigorous treatment
of the subject, following as much in parallel as possible with classical
information theory and, on the other side, stressing profound differences of
the quantum case. An emphasis is made on recent advances, and several still
unsolved problems are briefly outlined. 

 There is yet ``more quantum'' domain of problems concerning
reliable transmission of entire quantum states under a given fidelity
criterion \cite{xben}.  The very definition of the relevant ``quantum
information'' is far from obvious. Important steps in this direction were
made in \cite{xbarnum}, \cite{xdiv}, 
where in particular a tentative converse of the
relevant coding theorem was suggested.  However the proof of the
corresponding direct theorem remains an open question.
\vskip20pt
\centerline{\sc II.  General considerations}\vskip10pt
\centerline{\bf \S 1. The quantum communication channels}
\vskip10pt
A communication channel in general can be described as an affine mapping
which transforms states of the input system into states of the output system.
States represent statistical ensembles that can be mixed, and the affinity
property reflects fundamental requirement of preservation of the statistical
mixtures.  In case of classical systems states are described by probability
distributions, and classical communication channel is just a transition
mapping from input to output probability distributions. If at least one of
the systems is quantum, one speaks of quantum communication channel.

Let $\cal H$ be a Hilbert space providing a quantum-mechanical description
for the physical carrier of information. We do not ask $\cal H$ to be
finite-dimensional, as in quantum communication this may well be not the case
(while in applications to quantum computing finite dimensions always
suffice).  We shall not dwell upon topological questions (unless this is a
matter of principle as in \S IV.2), and the convergence of operator series
below is usually to be understood in the weak operator sense (although in
some cases it is in fact stronger, say in the norm sense).

A {\sl quantum state} is a {\sl density operator}, i. e.  positive operator
$S$ in $\cal H$ with unit trace, Tr$S = 1$.  Following Dirac's formalism, we
shall denote vectors of $\cal H$ as $ |\psi\rangle$, and hermitean conjugate
vectors of the dual space as $\langle\psi |$.  Then $\langle\phi |\psi
\rangle$ is the inner product of $|\phi\rangle, |\psi\rangle$
and $|\psi\rangle\langle\phi |$ is the outer product, i. e. operator $A$ of
rank 1, acting on vector $|\chi\rangle$ as $A |\chi\rangle =
|\psi\rangle\langle\phi |\chi\rangle.$ If $|\psi\rangle$ is a unit vector,
then $|\psi\rangle\langle\psi |$ is the orthogonal projection onto
$|\psi\rangle$. This is a special density operator, representing {\sl pure}
state of the system. Pure states are the extreme points of the convex set
${\cal S}({\cal H})$ of all states; an arbitrary state can be represented as
a mixture of pure states, i. e. by imposing classical randomness on pure
states.  In this sense pure states are ``noiseless'', i. e. they contain no
classical source of randomness. By the spectral theorem, every density
operator can be represented as a mixture of pure states $$S =
\sum_{i}\lambda_i |\psi_i\rangle\langle\psi_i |,$$
where $\lambda_i$ are the eigenvalues, $|\psi_i\rangle$ are the eigenvectors
of $S$.  Note that $\{\lambda_i\}$ form a probability distribution i.e. a
classical state on the set of eigenvectors of $S$. This also means that
classical states can always be embedded into ${\cal S}({\cal H})$ by fixing
some orthonormal system $\{|\psi_i\rangle\}$ in $\cal H$.

The following notion of {\sl quantum decision rule} is a far-reaching
generalization of the standard notion of observable. Mathematically it is
described by a {\sl resolution of identity} in $\cal H$, that is by a family
$X = \{ X_j\}$ of positive operators in $\cal H$ satisfying $\sum_j X_j = I$,
where $I$ is the unit operator in $\cal H$ .  The probability of taking a
decision $j$ if the decision rule $X$ is applied to system in the state $S$
is postulated by the following generalization of the Born statistical
formula: $$ P (j|S) = \mbox{Tr}S X_j .$$ From a physical point of view, a
decision rule is implemented by a quantum measurement including possible
posterior processing of the measurement results (see \cite{xprob},
\cite{xkraus} for more discussion).
The standard notion of observable is recovered if one requires $X_j$ to be
mutually orthogonal projection operators, $X_j X_k = \delta_{jk} X_j$.  The
mapping $S\rightarrow P(\cdot |S)$ is affine and it can be shown that any
affine mapping from quantum states to probability distributions has this form
(see \cite{xprob}, Proposition 1.6.1). In fact, it is already an example of
quantum channel ( q-c channel, see below).  A system $\{|\phi_j\rangle\}$ of
vectors in $\cal H$ is called {\sl overcomplete} if $\sum_j
|\phi_j\rangle\langle\phi_j | = I$.  Every overcomplete system (in particular
every orthonormal basis) gives rise to the decision rule $X$ for which $X_j =
|\phi_j\rangle\langle\phi_j |$ and $P(j|S) = \langle\phi_j |S \phi_j
\rangle.$

The classical case is embedded into this picture by assuming that all
operators in question commute, and hence are diagonal in some basis labelled
by index $\omega$; in fact by taking $S = \mbox{diag}[ S(\omega ) ], X_j =
\mbox{diag}[ X(j |\omega) ]$, we have the classical state $S$ and the classical
decision rule $X$, such that $P(j|S) = \sum_{\omega}X(j |\omega) S(\omega )$.
Standard quantum observables correspond then to classical deterministic
decision rules (random variables).

The earliest mathematical definitions of quantum communication channel
\cite{xechigo}
described it essentially as an affine mapping of the convex set ${\cal
S}({\cal H})$.  One sees easily that any such mapping $\Phi$ is a restriction
to the set of quantum states of a positive linear trace preserving mapping of
the space of trace-class operators, and vice versa.  However later it became
clear that such a definition should be substantially narrowed by imposing the
fundamental condition of complete positivity \cite{xhol72}, \cite{xkraus},
\cite{xlin}. A linear mapping
$\Phi$ is {\sl completely positive} if for any finite collections of vectors
$\{|\phi_i\rangle\}, \{|\psi_i\rangle\} \subset \cal H$
\begin{equation}\label{2.1}      \sum_{i,j} \langle\phi_i|\Phi [|\psi_i\rangle\langle\psi_j|]\phi_j\rangle\geq 0.\end{equation}
(this is only one of possible equivalent definitions). It turns out that this
property is necessary and sufficient for physical realizability of the
channel via unitary interaction with another quantum system (the environment)
\cite{xkraus}, \cite{xlin}. Basing on a fundamental result of Stinespring one shows that arbitrary linear completely positive trace preserving mapping can be written (non-uniquely) in the form
\begin{equation}\label{2.2}    \Phi [S] = \sum_{k} V_k S V_k^*  \end{equation}
with $\sum_{k}V_k^* V_k = I.$ We shall call any such mapping a {\sl channel}
\footnote{A recent paper \cite{xnagaoka} presents an attempt to investigate
the capacity of channels given by positive but not completely positive maps.
Such attempts may be interesting in view of recent observation \cite{xkos}
that such channels might be realizable via interactions with more
sophisticated environment (as non-Abelian gauge field) described by operators
in a graded tensor product of Hilbert spaces.}.

We now introduce an important class of channels. Let $\{ S_i\}$ be a family
of quantum states and $\{X_i\}$ a resolution of identity in $\cal H$. Let
\begin{equation}\label{2.3}     \Phi [S] = \sum_{i}  S_i\, \mbox{Tr}SX_i  .  \end{equation}
It is easy to check that this is linear completely positive trace preserving
mapping; it is a good exercise to find a representation (\ref{2.2}) for such
$\Phi$. If $X_i = |e_i\rangle\langle e_i|$, where $\{e_i\}$ is an orthonormal
basis, we call it {\sl classical-quantum (c-q) channel}. As easily seen, it
is equivalent to giving a mapping $i\rightarrow S_i$ from classical input
alphabet $A = \{i\}$ to quantum states.  If, moreover, all states $S_i$
commute the channel is called {\sl quasiclassical}; such channel is
equivalent to a classical channel with transition probability, given by the
eigenvalues $S(\omega |i)$ of the states $S_i$.

On the other hand, if $X_i$ are arbitrary and $S_i= |e_i\rangle\langle e_i|$,
we call the channel {\sl quantum-classical (q-c channel)}, as it is
equivalent to giving a decision rule that maps quantum states into
probability distributions on the output alphabet $B =\{i\}$. The channels of
the form (\ref{2.3}) by no means exhaust all possibilities; the simplest
example of a channel which is not of the form (\ref{2.3}) is given by
reversible evolution
\begin{equation}\label{2.4}     \Phi [S] =  V S V^* ,  \end{equation}
where $V$ is arbitrary unitary operator.
\vskip20pt
\centerline{\bf \S 2. The entropy bound and the capacity of quantum channel }
\vskip10pt
If $\pi$ is a discrete probability distribution on ${\cal S}({\cal H})$,
assigning probability $\pi_i$ to the state $S_i$, we denote
\begin{equation}\label{1-5}     \Delta H (\pi ) =
\sum_{i}\pi_i H(S_i ; {\bar S}_{\pi}) ,
\end{equation} where
\begin{equation}\label{1-6}     {\bar S}_{\pi} = \sum_{i}\pi_i S_i ,\end{equation}
and $$H(S ; S' ) = \mbox{Tr}S(\log S - \log S')$$ is quantum relative entropy
(see
\cite{xlin}, \cite{xwehrl}, \cite{xpetz} for more careful definition and discussion of the properties).
Just as the relative entropy, the quantity $ \Delta H (\pi )$ is nonnegative
but may be infinite. If \begin{equation}\label{1-7}
\sup_{i} H(S_i) < \infty ,
\end{equation}
where $H(S) = - \mbox{Tr}S\log S$ is the quantum entropy, then
\begin{equation} \label{1-8}
\Delta H (\pi ) = H({\bar S}_{\pi}) - {\bar H}( S_{(\cdot)}) ,\end{equation}
where ${\bar H}( S_{(\cdot)}) =\sum_{i}\pi_i H(S_i) < \infty.$

Let $X = \{X_j\}$ be a decision rule, and let $P(j|i) = \mbox{Tr}S_i X_j$.
We denote by
\begin{equation}\label{1-81}   I (\pi , X ) = \sum_j \sum_i \pi_i
P(j|i)\mbox{log}\left( \frac{P(j|i)} {\sum_k \pi_k P(j|k)}\right)
\end{equation}
the classical mutual information between input and output random variables.
The {\sl quantum entropy bound} says that
\begin{equation}\label{1-82}   \sup_{X}I (\pi, X)\leq \Delta H (\pi ),    \end{equation}
with the equality achieved if and only if all the operators $\pi_i S_i$
commute. The inequality was explicitly conjectured in \cite{xgordon} in the
context of conventional quantum measurement theory.  The proof for a finite
number of states in finite-dimensional Hilbert space based on the study of
convexity properties of the quantities in both sides of (\ref{1-82}) was
given in \cite{xhol73}.

A different approach to this bound is related to the strong subadditivity of
quantum entropy \cite{xlie} and equivalent property of decrease of
quantum relative entropy under trace preserving completely positive maps
developed later in the series of papers \cite{xlin} and in
\cite{xuhl}, namely
$$H(\Phi (S); \Phi (S')) \leq H(S; S')$$ for any states $S,S'$ and channel
$\Phi$.  This can be used to generalize the entropy bound to the case of
infinitely many states in infinite dimensions by choosing $\Phi$ to be the
q-c channel implying the quantum decision rule (cf. \cite{xyuen} ). It is
also not difficult to extend the initial proof given in \cite{xhol73}, but
the reformulation of the entropy bound in terms of the relative entropy is
important for a different reason: it extends to the case where (\ref{1-7})
does not hold, the signal states $S_i$ can have infinite entropy,
and the formula (\ref{1-8}) is no longer valid.

If $\Phi$ is a channel, we denote $I (\pi ,\Phi, X )$ the mutual information
defined analogously to (\ref{1-81}), but with the transition probabilities
given by $P(j|i) = \mbox{Tr}\Phi[S_i] X_j$, and
\begin{equation}\label{1-51}     \Delta H (\pi, \Phi ) =
\sum_{i}\pi_i H(\Phi [S_i] ; \Phi [{\bar S}_{\pi}]) .
\end{equation} In order to consider block codes let us introduce the product
channel ${\Phi}^{\otimes n} = {\Phi}\otimes ...\otimes {\Phi}$ in the Hilbert
space ${\cal H}^{\otimes n} = {\cal H}\otimes ...\otimes {\cal H}$ .  Let us
denote
\begin{equation}\label{1-52}   C_n (\Phi ) =  \sup_{\pi} \sup_{X}I (\pi ,{\Phi}
^{\otimes n}, X );
\quad{\bar C}_n (\Phi ) =  \sup_{\pi} \Delta H(\pi, {\Phi}
^{\otimes n} ), \end{equation} where the suprema are taken over all discrete
probability distributions $\pi$ on ${\cal S}({\cal H} ^{\otimes n})$, and
over all decision rules $X$ in ${\cal H} ^{\otimes n}$. It is easily seen, by
taking product probability distributions $\pi$, that the quantities $C_n
(\Phi ), {\bar C}_n (\Phi )$ are superadditive: $$ C_n (\Phi )+ C_m (\Phi )
\leq  C_{n+m} (\Phi ),
\qquad {\bar C}_n (\Phi )+ {\bar C}_m (\Phi ) \leq{\bar C}_{n+m} (\Phi ).$$
This implies that the following limits exist
\begin{equation}\label{1-53}      C (\Phi ) = \lim_{n \to \infty} C_n (\Phi )/n =
\sup_{n} C_n (\Phi )/n , \end{equation}\begin{equation}\label{1-54}
{\bar C} (\Phi ) = \lim_{n \to \infty}{\bar C}_n (\Phi )/n =
\sup_{n}{\bar C}_n (\Phi )/n .
\end{equation} The entropy bound implies $$ C (\Phi ) \leq {\bar C} (\Phi
).$$ We call the quantity $ C (\Phi )$ the {\sl capacity} of the channel
$\Phi$. This definition is naturally justified by an application of the
classical Shannon coding theorem (cf. \cite{xhol79}), but we shall give a
different argument implying also (under some regularity conditions) much
stronger statement
\begin{equation}\label{1-55}  C (\Phi ) = {\bar C} (\Phi ) . \end{equation}

For a classical channel $C_n (\Phi )= n C_1(\Phi )$ is additive, and
trivially $ C (\Phi ) = {\bar C} (\Phi ) = C_1 (\Phi )$. A striking feature
of quantum case is possibility of the inequality $C_1 (\Phi )< C (\Phi )$
implying strict superadditivity of the information quantities $C_n (\Phi )$
(see \S III.2,3). In a sense, there is a kind of ``quantum memory'' in
channels, which are the analog of classical memoryless channels. This fact is
just another manifestation of the ``quantum nonseparability'', and in a sense
is dual to the existence of Einstein - Podolsky - Rosen correlations: the
latter are due to entangled (non-factorizable) states and hold for
disentangled measurements while the superadditivity is due to entangled
measurements and holds for disentangled states \cite{xhol79}, \cite{xperez}.

The paper \cite{xbennet} raised the general question of (super)additivity of
the quantities ${\bar C}_n (\Phi )$.  If they are additive then $${\bar C}
(\Phi ) ={\bar C}_1 (\Phi ) = \sup_{\pi}
\Delta H( \pi, {\Phi})$$
which further greatly simplifies calculation of the capacity. This trivially
holds for reversible channels (\ref{2.4}). The following Proposition shows
that this is also true in somewhat opposite cases.

{\bf Proposition 1}. {\sl If $\Phi$ is c-q or q-c channel, then $$ {\bar C}_n
(\Phi )+ {\bar C}_m (\Phi ) = {\bar C}_{n+m} (\Phi ).$$

Proof. } It is sufficient to show that
\begin{equation}\label{1-56}
{\bar C}_1 (\Phi )+ {\bar C}_1 (\Phi ) \geq {\bar C}_{2} (\Phi ).
\end{equation}

If $\Phi$ is a c-q channel,
\begin{equation}\label{1-57}     \Phi [S] = \sum_{i}  S_i \,\langle e_i|S e_i\rangle,  \end{equation}
where $S_i$ are fixed states in ${\cal H}$, then $$ \sup_{\pi}\Delta H (\pi,
\Phi  ) =    \sup_{\pi_i} \Delta H ( \pi ) ,$$
where $ \Delta H ( \pi ) $ is given by the expression (\ref{1-5}) with these
fixed states $S_i$. Let the distribution $\pi$ assign the probability
$\pi_{ij}$ to the state $S_i\otimes S_j$ in ${\cal H}\otimes{\cal H}$. We
have
\begin{equation}\label{1-58}     \Delta H ( \pi ) \leq  \Delta H ( \pi^1 )  + \Delta H ( \pi^2 ) ,  \end{equation}
where $\pi^1$ is the first marginal distribution of $\pi$ assigning
probability $\pi_i^1=\sum_{j}\pi_{ij}$ to the state $S_i$ in ${\cal H}$, and
similarly $\pi^2$ is the second marginal distribution assigning probability
$\pi_j^2=\sum_{i}\pi_{ij}$ to the state $S_j$ in ${\cal H}$.  In finite
dimensional case where formula (\ref{1-8}) always holds, this follows from
subadditivity of the entropy with respect to tensor products \cite{xwehrl}
(see the proof of Lemma 2 in Appendix of \cite{xhol}). In infinite
dimensional case let us consider a monotonously increasing sequence of
orthogonal projections $P_r\uparrow I$ in $\cal H$, and introduce $$ \Delta
H_r (\pi ) =
\sum_{i}\pi_i H(P_r S_i P_r; P_r{\bar S}_{\pi}P_r).$$ By the properties of relative entropy \cite{xlin}, $ \Delta H_r (\pi )\uparrow  \Delta H (\pi )$. By using (\ref{1-58}) for normalized projected states, we obtain
$$ \Delta_r H ( \pi ) \leq \Delta H_r ( \pi^1 ) + \Delta H_r ( \pi^2 ) - \phi
(
\mbox{Tr} (P_r\otimes P_r){\bar S}_{\pi}(P_r\otimes P_r)),$$ where $\phi (x) = - x\log x$.
Passing to the limit $r\rightarrow \infty$ gives (\ref{1-58}) in the general
case. Taking in (\ref{1-58}) supremum over $\pi$ gives (\ref{1-56}).

Now let $\Phi$ be a q-c channel
\begin{equation}\label{1-59}     \Phi [S] = \sum_{j}  \mbox{Tr}S X_j \,| e_j\rangle\langle e_j|,  \end{equation}
and let $\pi$ be a discrete probability distribution on ${\cal S}({\cal H})$,
assigning probability $\pi_k$ to a state $S_k$, then the density operators $
\Phi [S_k] $ commute
and $$ \Delta H (\pi, \Phi ) = I ({\cal K};{\cal J}), $$ is the classical
mutual information (\ref{1-81}) corresponding to the input probability
distribution $\pi$ and transition probability $P(j|k)=\mbox{Tr} S_k X_j$.
Here we denote by $\cal K$ the input random variable taking values $k$, and
by $\cal J$ the output random variable taking the values $j$. In order to
prove (\ref{1-56}), consider states $S_k$ in the Hilbert space ${\cal
H}\otimes {\cal H}$, then the transition probability is
\begin{equation}\label{1-60}    P(j_1,j_2|k) = \mbox{Tr} S_k (X_{j_1}\otimes  X_{j_2}) = P^1(j_1|j_2,k) P^2(j_2|k) ,  \end{equation}
where $$P^1(j_1|j_2,k) = \mbox{Tr}_1 S^1_{j_2,k} X_{j_1}, P^2(j_2|k) =
\mbox{Tr}_2 S^2_k X_{j_2},$$ and
$$S^2_k = \mbox{Tr}_1 S_k ,\qquad S^1_{j_2,k} = { \mbox{Tr}_2 S_k (I\otimes
X_{j_2}) \over \mbox{Tr} S_k (I\otimes X_{j_2})}.$$ Here we denote by $
\mbox{Tr}_r$
(partial) trace with respect to $r$-th factor ($r=1,2$) in ${\cal
H}\otimes{\cal H}$.

We then have $$ \Delta H (\pi, \Phi\otimes\Phi ) = I ({\cal K};{\cal
J}_1{\cal J}_2)= H({\cal J}_1{\cal J}_2) - H({\cal J}_1{\cal J}_2| {\cal K}),
$$ where $H(\cdot), H(\cdot|\cdot)$ are, respectively, classical entropy and
conditional entropy of the random variables. By subadditivity of the
classical entropy, $$H({\cal J}_1{\cal J}_2) \leq H({\cal J}_1) + H({\cal
J}_2).$$ On the other hand, (\ref{1-60}) implies $$H({\cal J}_1{\cal J}_2|
{\cal K}) = H({\cal J}_1|{\cal J}_2 {\cal K}) + H({\cal J}_2| {\cal K}).$$
Combining, we get $$ I ({\cal K};{\cal J}_1{\cal J}_2)\leq I ({\cal K}{\cal
J}_2; {\cal J}_1) + I ({\cal K};{\cal J}_2), $$ which amounts to $$ \Delta H
(\pi, \Phi\otimes\Phi )\leq \Delta H ( \pi ^1, \Phi ) +
\Delta H ({ \pi}^2,\Phi ),$$ where ${ \pi}^1$ is the probability distribution on ${\cal S}({\cal H})$,
assigning probability $\pi_k P^2(j_2 |k)$ to the state $ S^1_{j_2,k}$ and ${
\pi}^2$ is the probability distribution on ${\cal S}({\cal H})$,
assigning probability $\pi_k $ to the state $ S_k^2$.  Taking supremum over
$\pi$ gives (\ref{1-56}). $\Box$
\newpage
\centerline{\bf \S 3. Formulation of the quantum coding theorem. The weak converse }
\vskip10pt
We call by {\sl code of size} $M$ a sequence $(S^1 , X_1 ),..., (S^M , X_M
)$, where $S^k$ are states, and $\{ X_k \}$ is a family of positive operators
in ${\cal H}^{\otimes n}$, satisfying $\sum_{k=1}^M X_k \leq I$. Defining
$X_0 = I - \sum_k X_k$, we have a resolution of identity in ${\cal
H}^{\otimes n}$.  An output $k \geq 1$ means decision that the state $S^k$
was transmitted, while the output $0$ is interpreted as evasion of any
decision.The average error probability for such a code is
\begin{equation} \label{1-4}{\bar \lambda} (S, X) =
\frac{1}{M} \sum_{k=1}^M [1 - \mbox{Tr}\Phi[S^k ] X_k ]. \end{equation}
Let us denote $p(n, M)$ the infimum of this error probability with respect to
all codes of the size $M$.

{\bf Theorem 1.} {\sl If $C (\Phi )< \infty$ and $R > C(\Phi )$ , then $ p(n,
\mbox{e}^{nR})\not\rightarrow 0$.  On the other hand, let $\Phi$ be a channel satisfying the condition
\begin{equation}\label{1-61}
\sup_{S\in {\cal S}({\cal H})} H(\Phi [S]) < \infty,
\end{equation} then $ p(n, \mbox{e}^{nR})\rightarrow 0$ for $R < {\bar
C}(\Phi )$. In particular, $C(\Phi ) = {\bar C}(\Phi )$.

Proof}. The proof of the first statement is based on the inequality
\begin{equation}\label{1-10}
\log M \cdot(1- p(n, M))\leq  C_n (\Phi ) + 1,
\end{equation} which is simple corollary of the classical Fano inequality.
Indeed, let $\cal J$ be the classical random variable describing the output
of the product channel under the decision rule $X$ if the words in the code
$(S, X)$ are taken with the input distribution $\pi_M$ assigning equal
probability $1/M$ to each state $S^k$ , and let $\cal K$ be the random
variable, the value of which is the number $k$ of the transmitted state.  The
Fano inequality \cite{xgal}, \cite{xinf} implies $$\log M \cdot (1-{\bar
\lambda} (S, X))\leq I({\cal K}; {\cal J}) +1\leq C_n (\Phi ) +1.$$
Taking $M = \mbox{e}^{nR}$ and letting $n\rightarrow\infty$, we come to the
conclusion $ p(n, \mbox{e}^{nR})\not\rightarrow 0$.

As for the second statement, here we shall show only how the proof for the
general case reduces to the case of c-q channel (\ref{1-57}) satisfying the
condition
\begin{equation}\label{1-62}     \sup_i H(S_i) < \infty. \end{equation} The following Chapter will be devoted essentially to the treatment of that special case.

If $R < {\bar C}(\Phi )$, then we can choose $n_0$ and probability
distribution $\pi^0$ on ${\cal S}({\cal H}^{\otimes n_0})$ such that $n_0 R <
\Delta H ( \pi^0, \Phi^{\otimes n_0} ).$ Let $\pi^0$  assign probability $\pi_i^0$ to the state $S_i$ in
${\cal H}^{\otimes n_0}$, and consider the c-q channel $\tilde \Phi$ in this
Hilbert space given by the formula $${\tilde \Phi}[S] = \sum_{i}\Phi^{\otimes
n_0}[S_i] \,\langle e_i|S e_i\rangle.$$ According to Proposition 1, $$ {\bar
C}(\tilde \Phi ) = \sup_{\pi_i} \left\{H(\sum_{i}\pi_i \Phi^{\otimes
n_0}[S_i]) -
\sum_{i}\pi_i H(\Phi^{\otimes n_0}[S_i])\right\},$$
which is greater than $n_0 R$.  Denoting ${\tilde p}({ n}, { M})$ the minimal
error probability for $\tilde \Phi$, we have (assuming $n$ to be multiple of
$n_0$) $$p(n, \mbox{e}^{nR}) \leq {\tilde p}({n/ n_0},
\mbox{e}^{(n/n_0)n_0R}),$$ since every code of size $M$ for  $\tilde \Phi$
is also code of the same size for $ \Phi$.  It follows that if we prove the
statement for the c-q channel $\tilde \Phi$, it will be also proved for the
initial channel $\Phi$.

Let us now show that the condition (\ref{1-61}) implies (\ref{1-62}) for the
channel $\tilde \Phi$.  Indeed, $$\sup_{i}H (\Phi^{\otimes n_0}[S_i]) \leq
\sup_{S\in{\cal S}({\cal H}^{\otimes n_0})}H(\Phi^{\otimes n_0}[S]) \leq n_0 \sup_{S\in{\cal S}({\cal H})}H(\Phi[S]) < \infty,$$
by subadditivity of quantum entropy with respect to tensor products.$\Box$

From now on we shall consider c-q channel (\ref{1-57}) in the Hilbert space
$\cal H$, determined by the mapping $i\rightarrow S_i$ from the input
alphabet $A=\{i\}$ to ${\cal S}({\cal H})$, and shall skip $ \Phi$ from all
notations.  For c-q channel the output states are fixed, and sending a word $w
= (i_{1},\ldots,i_{n})$ produces the tensor product state $S_w =
S_{i_1}\otimes \ldots \otimes S_{i_n}$.  A code of size $M$ is a collection
$(w^1, X_1), ... , (w^M, X_M)$, where $w^k$ are words of lengths $n$.  The
average error probability of the code is
\begin{equation}\label{1-63}    \bar{\lambda}(W,X) =
{1 \over M} \sum_{k=1}^{M} [1 - \mbox{Tr}S_{w^k}X_{k}] \end{equation}

In terms of our previous definition this means that the input states can be
taken as products of the pure states
$$S^k = |e_{i_1}\rangle\langle e_{i_1}|\otimes\dots\otimes|e_{i_n}\rangle
\langle e_{i_n}|,$$ where $|e_{i}\rangle$ are taken from the representation
(\ref{1-57}). Using more general input states $S^k$ amounts to randomly
chosen codewords, which cannot increase the rate of information transmission.
The proof of the Theorem will be completed in \S III.2, but first we discuss
in more detail pure state channels.
\vskip20pt\centerline{\sc III. The proof of the direct quantum coding theorem}\vskip10pt
\centerline{\bf \S 1. The pure state channels}
\vskip10pt
Let us consider a {\sl pure state channel} with $S_i = |\psi_i
\rangle\langle\psi_i|$ . Since the entropy of a pure state is
zero, the condition (\ref{1-7}) is trivially satisfied and $\Delta H (\pi ) =
H( {\bar S}_{\pi})$ for such a channel. By discussing pure state channel
first we shall follow historical development of the subject and prepare for
considerably more technical treatment of the general c-q channel.
Also in this case we can obtain more advanced results concerning the
asymptotic behavior of the error probability and the reliability function
that are still unavailable in the general case.

For a pure state channel sending a word $w = (i_{1},\ldots,i_{n})$ produces
the tensor product vector $\,\psi_{w} =
\psi_{i_{1}} \otimes \ldots \otimes \psi_{i_{n}} \in {\cal H}^{\otimes n}$.
We are now interested in obtaining upper bounds for the error probability
$p(n, M)$ minimized over all codes of size $M$.The first step has geometric
nature and amounts to obtaining a tractable upper bound for the average error
probability
\begin{equation}\label{V1}    \bar{\lambda}(W,X) =
{1 \over M} \sum_{k=1}^{M} [1 - \langle\psi_{w^k}|X_{k}\psi_{w^k}\rangle ]
\end{equation}
minimized over all decision rules.  Minimization of (\ref{V1}) is the quantum
Bayes problem with uniform apriori distribution, and its solution is a
natural analog of the maximal likelihood decision rule.  There are necessary
and sufficient conditions for solution of the quantum Bayes problem
\cite{xhol74},  which, however, can be solved explicitly only in some particular cases,
especially if the family of states has certain symmetry.  It is therefore
necessary to look for a suitable approximation of the quantum maximum
likelihood decision rule.

Let us restrict for a while to the subspace of ${\cal H}^{\otimes n}$
generated by the code vectors $\psi_{w^1},\ldots,\psi_{w^M}$, and consider
the Gram matrix $\,\Gamma = [\langle\psi_{w^i}|\psi_{w^j}\rangle]\,$ and the
Gram operator $G =
\sum_{k=1}^{M}|\psi_{w^k}\rangle\langle\psi_{w^k}|$.
This operator has the matrix $\Gamma $ with respect to the overcomplete
system
\begin{equation}\label{iii1}
|\hat{\psi}_{w^k}\rangle = G^{-1/2} |{\psi}_{w^k}\rangle\,;\,
\quad k=1,\ldots,M\;. \end{equation}
Following \cite{xhol78} consider the resolution of identity
\begin{equation}\label{iii2}    X_{k} = |\hat{\psi}_{k}\rangle\langle\hat{\psi}_{k}|  \end{equation} which will approximate
the quantum maximum likelihood decision rule ; the necessary normalizing
factor $G^{-1/2}$ is the source of entanglement in the decision rule (it is
also a major source of analytical difficulties in the noncommutative case).
Note that the vectors $\psi_{w^1},\ldots,\psi_{w^M}$ need not be linearly
independent; in the case of linearly independent coherent state vectors
(\ref{iii2}) is related to the ``suboptimal receiver'' described in
\cite{xhel}, Sec. VI.3(e).
By using this decision rule we obtain the upper bound
\begin{equation}\label{iii3}
\inf_{X} \bar{\lambda}(W, X) \leq {2 \over M}\,{\rm Sp}
\left(E - \Gamma ^{1/2}\right) = {1 \over M}\,{\rm Sp}
\left(E - \Gamma ^{1/2}\right)^2 ,
\end{equation} where $E$ is the unit $M\times M$-matrix and $\rm Sp$ is the
trace of $M\times M$-matrix. Indeed, for the decision rule
(\ref{iii2})$$\bar{\lambda}(W, X) = {1 \over M}\sum_{k=1}^{M} [1 -
|\langle\psi_{w^k}|{\hat \psi}_{w^k}\rangle|^2 ]$$ $$\leq {2 \over M}
\sum_{k=1}^{M}
[1 - |\langle\psi_{w^k}|{\hat \psi}_{w^k}\rangle|] = {2 \over M}
\sum_{k=1}^{M}
[1 - \langle{\hat \psi_{w^k}}| G^{1 \over 2} {\hat \psi}_{w^k}\rangle ],$$
which is (\ref{iii3}).  In deriving second relation in (\ref{iii3}) we used
${\rm Sp} \Gamma = {\rm Sp} E = M.$
This bound is ``tight'' in the sense that there is a similar lower bound
\cite{xhol78}.
However it is difficult to use because of the presence of square root of the
Gram matrix. A simpler but coarser bound is obtained by using the 
inequality 
\begin{equation}\label{ine}(1 - x^{1/2})^2 = (1 - x )^2 (1 +
x^{1/2})^{-2}  \leq (1 - x )^2, \quad x\geq 0,\end{equation}
 applied to eigenvalues of
$\Gamma$:
\begin{equation}\label{iii4}
\inf_{X} \bar{\lambda}(W, X) \leq {1 \over M}\,{\rm Sp}
\left(E - \Gamma \right)^2 = {1 \over M}\,{\rm Tr}
\sum\sum_{r\not= s} S_{w^r} S_{w^s}.
\end{equation} As shown in \cite{xhol78}, this bound is asymptotically
equivalent (up to the factor 1/4) to the tight bound (\ref{iii3}) in the
limit of ``almost orthogonal'' states $\Gamma \rightarrow E$. On the other
hand, different words are ``decoupled'' in (\ref{iii4}) which makes it
suitable for application of the random coding.

Just as in the classical case, we assume that the words $w^1 ,..., w^M$ are
chosen at random, independently and with the probability distribution
\begin{equation}\label{iii5}      {\sf P}\{w =( i_1,\ldots,i_n)\} = \pi_1 \cdot\ldots\cdot \pi_n . \end{equation}
Then for each word $w$ the expectation
\begin{equation}\label{iii6}    {\sf E}\,  S_w = \sum_{i_1,\ldots,i_n}
\pi_{i^1} \cdot\ldots\cdot \pi_{i^n} |\psi_{i_1}\rangle\langle \psi_{i_1}|\otimes\ldots\otimes
|\psi_{i_n}\rangle\langle \psi_{i_n}| = {\bar S}_{\pi}^{\otimes n},
\end{equation}
and by taking the expectation of the coarse bound (\ref{iii4}) we obtain, due
to the independence of $w^r, w^s$ $$p ( n, M)\leq {\sf E}
\inf_{X}\bar{\lambda}(W, X) \leq (M-1) {\rm Tr}
({\bar S}_{\pi}^{\otimes n})^2 = (M-1)\mbox{e}^{-n\log {\rm Tr}{\bar
S}_{\pi}^2}.$$ By denoting \begin{equation}\label{iii7} {\tilde C} =
-\log\inf_{\pi}{\rm Tr}{\bar S}_{\pi}^2 = -\log\inf_{\pi}\sum_{i,j}\pi_i\pi_j
|\langle\psi_i |\psi_j \rangle|^2, \end{equation} we conclude that $C\geq
{\tilde C}$ . There are cases (e. g.  pure state binary channel, see below)
where ${\tilde C} > C_1$, so this suffices to establish possibility of the
inequality $C > C_1$, and hence the strict superadditivity of $C_n$
\cite{xhol79}, but not sufficient
to prove the coding theorem, since ${\tilde C} <{\bar C}$ unless the channel
is quasiclassical. A detailed comparison of the quantities $C_1 , {\bar C}$
for different quantum channels was made by Ban, Hirota, Kato, Osaki and
Suzuki \cite{xban96}. The quantity $\tilde C$ was discussed in
\cite{xhol79}, \cite{xstr}, but its real information theoretic meaning
is elucidated only in connection with reliability function and 
quantum ``cutoff rate'' (see
\cite{xban98}).

The proof of the inequality $C\geq {\bar C}$ given in \cite{xjozsa} achieves
the goal by using the approximate maximum likelihood improved with projection
onto the ``typical subspace'' of the density operator ${\bar
S}_{\pi}^{\otimes n}$ and the correspondingly modified coarse bound for the
error probability. The coarseness of the bound is thus compensated by
eliminating ``non-typical'' (and hence far from being orthogonal) components
of the signal state vectors. More precisely, let us fix small positive
$\delta$, and let $\lambda_j$ be the eigenvalues, $|e_j \rangle$ the
eigenvectors of ${\bar S}_{\pi}$.  Then the eigenvalues and eigenvectors of
${\bar S}_{\pi}^{\otimes n}$ are $\lambda_J = \lambda_{j_1} \cdot ... \cdot
\lambda_{j_n},\quad |e_J \rangle = |e_{j_1}\rangle\otimes ...\otimes |e_{j_n}\rangle$
where $J = (j_1,...,j_n )$. The spectral projector onto the {\sl typical
subspace} is defined as $$P= \sum_{J\in B} |e_J \rangle\langle e_J |, $$
where $B = \{J: \mbox{e}^{-n[H({\bar S}_{\pi})+\delta ]} < \lambda_J <
\mbox{e}^{-n[H({\bar S}_{\pi})-\delta ]}\}$. This concept plays a central role
in ``quantum data compression'' \cite{xschum}. In a more mathematical context
a similar notion appeared in \cite{xpetz}, Theorem 1.18. Its application to
the present problem relies upon the following two basic properties: first, by
definition,
\begin{equation}\label{iii8}     \|{\bar S}_{\pi}^{\otimes n} P\| < \mbox{e}^{-n[H({\bar S}_{\pi})-\delta ]}.
\end{equation} Second, for fixed small positive $\epsilon$ and large enough
$n$
\begin{equation}\label{iii9}     \mbox{Tr}{\bar S}_{\pi}^{\otimes n}(I - P)\leq\epsilon ,    \end{equation}
because a sequence $J\in B$ is typical for the probability distribution given
by eigenvalues $\lambda_J$ in the sense of classical information theory
\cite{xgal}, \cite{xinf}.

By replacing the signal state vectors $|\psi_{w^k}\rangle$ with unnormalized
vectors $|{\tilde \psi}_{w^k}\rangle=P|\psi_{w^k}\rangle$, defining the
corresponding approximate maximum likelihood decision rule,and denoting
${\tilde \Gamma }$ the corresponding Gram matrix, the upper bound
(\ref{iii3}) is modified to $$
\inf_{X} \bar{\lambda}(W, X) \leq {2 \over M}\,{\rm Sp}
\left(E - {\tilde \Gamma}^{1/2}\right).$$
By using the inequality
\begin{equation}\label{ner} 
2-2x^{1/2}=(1-x)+(1-x^{1/2})^{2}\leq
(1-x)+(1-x)^{2},\quad x\geq 0,
\end{equation}
which follows from (\ref{ine}), we can obtain
$$
\inf_{X} \bar{\lambda}(W, X) \leq {1 \over M}\,\{{\rm Sp}
\left(E - {\tilde \Gamma} \right) +
{\rm Sp}
\left(E - {\tilde \Gamma} \right)^2\}$$  $$
\leq  {1 \over M}\,\sum_k \{ 2 {\rm Tr} S_{w^k}(I-P) +
\sum_{k\not= l} {\rm Tr}S_{w^k} P S_{w^l} P\}.$$
 Applying the random coding and using
(\ref{iii6}) and the properties (\ref{iii8}), (\ref{iii9}) of the typical
subspace, one gets for large $n$ $$p( n, M)\leq 2\mbox{Tr}{\bar
S}_{\pi}^{\otimes n}(I - P) + \mbox{Tr}({\bar S}_{\pi}^{\otimes n}P)^2 \leq
2\epsilon + (M-1)\mbox{e}^{-n[H({\bar S}_{\pi}) - \delta ]},$$ resulting in
$p( n, \mbox{e}^{nR})\rightarrow 0$ for $R<\bar{C}-\delta$, and hence in the
inequality $C\geq {\bar C}$.
\vskip20pt
\centerline{\bf \S 2. The quantum reliability function}
\vskip10pt
In classical information theory the coding theorem can be proved
without resorting to typical sequences, by mere use of clever estimates
for the error probability \cite{xgal}. Moreover, in this way one obtains
the exponential rate of convergence for the error probability, the
so called reliability function
\begin{equation}\label{rel}     E(R) = \lim_{n \to \infty}
\sup {1 \over n}\log {1 \over p(n, \mbox{e}^{nR})} \;,\quad 0 < R < C \;. \end{equation}
This puts us onto the idea of  trying to obtain similar estimates
 in the quantum case.

{\bf Theorem 2}. {\sl For all $M,n$ and $0\leq s\leq 1$
\begin{equation}\label{t2}      {\sf E}\,\inf_{X}\bar{\lambda}(W, X) \leq
2(M-1)^{s}\left[{\rm Tr}\,{\bar S}_{\pi}^{1+s} \right]^{n}. \end{equation}

Proof}. The first step of our argument is to remark that
\begin{equation} \label{f7}
{\frac{2}{M}}\,{\rm Sp(}E-\Gamma ^{1/2})={\frac{2}{M}}%
(M-{\rm Tr}G^{1/2}).
\end{equation}
 Consider
two operator inequalities
\[
-2G^{1/2}\leq -2G+2G,
\]
\[
-2G^{1/2}\leq -2G+(G^{2}-G).
\]
The first one is obvious, while the second follows from (\ref{ner}).
 Taking the expectation with respect to the probability
distribution  (\ref{iii5}), we get
\[
-2\,{\sf E}G^{1/2}\leq -2{\sf E}G+\left\{ 
\begin{array}{l}
2\,{\sf E}G \\ 
{\sf E}(G^{2}-G)
\end{array}
\right. .
\]
By using (\ref{iii6}), we obtain
\[
{\sf E}G={\sf E}\sum_{k=1}^{M}|\psi _{u^{k}}><\psi _{u^{k}}|=M{\sf E}|\psi
_{u^{k}}><\psi _{u^{k}}|=M{\bar{S}}_{\pi }^{\otimes n},
\]
\[
{\sf E}(G^{2}-G)={\sf E}\sum_{k=1}^{M}\sum_{l=1}^{M}|\psi _{u^{k}}><\psi
_{u^{k}}||\psi _{u^{l}}><\psi _{u^{l}}|-{\sf E}\sum_{k=1}^{M}|\psi
_{u^{k}}><\psi _{u^{k}}|\] \[ ={\sf E}\sum_{k\neq l}|\psi _{u^{k}}><\psi
_{u^{k}}|\psi _{u^{l}}><\psi _{u^{l}}|
=M(M-1)\left[ {\bar{S}}_{\pi }^{\otimes n}\right] ^{2}.
\]
Let$\left\{ e_{J}\right\} $ be the orthonormal basis of eigenvectors, and $%
\lambda _{J}$ the corresponding eigenvalues of the operator ${\bar{S}}_{\pi
}^{\otimes n}.$ Then
\[
-2\left\langle e_{J}|{\sf E}G^{1/2}|e_{J}\right\rangle \leq
 -2M\lambda _{J}+M\lambda _{J}\min
\left( 2,(M-1)\lambda _{J}\right) .
\]
By using the inequality $\,\min \{a,b\}\leq a^{s}b^{1-s},\,0\leq s\leq 1,$ we
get
 \[
\min \left( 2,(M-1)\lambda _{J}\right) \leq 2(M-1)^{s}\lambda
_{J}^{s}\,,\;0\leq s\leq 1\,.
\]
Summing with respect to $J$ and dividing by $M,$ we get from  (\ref{iii3}), (%
\ref{f7})
\[
{\sf E}\inf_X \bar{\lambda}(W, X)\leq 2(M-1)^{s}\sum_{J}\lambda
_{J}^{1+s}=2(M-1)^{s}\left[ {\rm Tr}\,{\bar{S}}_{\pi }^{1+s}\right]
^{n}\,,\;0\leq s\leq 1. \qquad \Box
\]

It is natural to introduce the function $\mu(\pi,s)$ similar to
analogous function in classical information theory
(\cite{xgal}, Ch. 5)
\begin{equation} \label{f19}
\mu (\pi,s) = 
- \log {\rm Tr}\,{\bar S}_{\pi}^{1+s}.
\end{equation}
By taking $M= \mbox{e}^{nR}$, we obtain
\begin{equation} \label{f20}
E(R) \geq
 \sup_{0\leq s \leq 1} \left(\sup_{\pi}\mu(\pi,s)-s R\right)\,
\equiv E_r (R).
\end{equation}
In particular, it follows easily that $$C\geq \sup_{\pi}\mu'(\pi , 0) = {\bar C}.$$
Thus the rate $C-\delta$ can be attained with the approximate 
maximum likelihood decision
rule (\ref{iii2}), (\ref{iii1}) without even projecting onto the typical subspace.

On the other hand, it appears possible to apply in the quantum case the
``expurgation'' technique from \cite{xgal},
Sec. 5.7, resulting in the bound
$$ E(R)\geq\sup_{s \geq 1} ( \sup_{\pi}{\tilde \mu}(\pi, s) -
s R )\equiv E_{ex}(R) ,$$ where
$${\tilde \mu}(\pi, s) = - s \log
\sum_{i,k }\pi_i \pi_k |\langle\psi_i |\psi_k\rangle|^{2 \over s}.$$
The behavior of the lower bounds $E_r (R), E_{ex} (R)$ can be studied by
the methods of classical information theory, see \cite{xbur}, and  is indeed similar
to the classical picture, where  $E_r (R)$ gives better bound for big rates $R$,
while $ E_{ex} (R)$ is better for small rates; in an intermediate region 
of rates the bounds $E_r (R), E_{ex} (R)$ have common linear portion $\tilde{C} - R$,
where
\begin{equation}\label{iv}   \tilde{C} = \sup_{\pi} {\tilde \mu}(\pi, 1) = \sup_{\pi}\mu (\pi, 1) = - \log\inf_{\pi}\mbox{Tr}{\bar
S}_{\pi}^2. \end{equation} This 
means that the quantity (\ref{iii7}) is a quantum analog of the ``cutoff
rate'' \cite{xban98}, a concept widely used in practical applications of
information theory. Figures 1, 2 present typical behavior of the functions
$\mu, {\tilde \mu}$ and Gallager's exponents $E_r, E_{ex}$ (modeled from the
binary quantum channel, see below).
\vskip20pt
\centerline{\bf \S 3. The binary quantum channel} 
\vskip10pt
Maximization of the bounds $E_{r}(\pi, R), E_{ ex}(\pi, R)$
over $\pi$, which is a difficult problem even in the classical case,
is still more difficult in quantum case. However, if the distribution
${\pi}^0$ maximizing either $\mu (\pi, s)$ or ${\tilde \mu}(\pi, s)$
is the same for all $s$, then $$E_r (R) = E_r ({\pi}^0, R),\qquad E_{ex} (R) = 
E_{ ex} ({\pi}^0, R).$$
This is the case for the binary pure state channel.

Let  $\,|\psi_0\rangle,\, |\psi_1\rangle$ be two
pure state vectors with $|\langle\psi_0|\psi_1\rangle| = \varepsilon $, in two dimensional Hilbert space $\cal H$. Consider the
operator $\,S_{\pi} = (1-\pi) S_{0} + \pi S_{1}\,$, where in notations the distribution 
$\pi$ is identified with the probability of the letter 1. Its
eigenvectors have the form $|\psi_0\rangle + \alpha |\psi_1\rangle$ with some
$\alpha$. Therefore for its eigenvalues we get the equation
$$
\left((1-\pi) |\psi_0\rangle\langle\psi_0| + \pi |\psi_1\rangle\langle\psi_1|\right)
\left(|\psi_0\rangle + \alpha |\psi_1\rangle\right) =
\lambda \left(|\psi_0\rangle + \alpha |\psi_1\rangle\right).
$$
Solving it, we find the eigenvalues
$$
\lambda_{1}(\pi) = {1 \over 2} \left[1 -
\sqrt{1 - 4(1-\varepsilon^{2})\pi (1-\pi)}\right] \;,
$$
$$
\lambda_{2}(\pi) = {1 \over 2} \left[1 +
\sqrt{1 - 4(1-\varepsilon^{2})\pi (1-\pi)}\right] \;.
$$

It is easy to check that both functions
$$\mu (\pi, s) = - \log \left(\lambda_1 (\pi )^{1+s} + \lambda_2 
(\pi )^{1+s}\right),$$ $${\tilde \mu}(\pi, s) = - s \log\left(
\pi^2 + (1-\pi )^2 + 2\pi (1 - \pi )\varepsilon^{2/s}\right)$$
are maximized by $\pi = 1/2$. Denoting $$\mu (s) = \mu(1/2, s) =
- \log \left[\left({1-\varepsilon \over 2}\right)^{1+s} + 
\left({1+\varepsilon \over 2}\right)^{1+s}\right],$$
$${\tilde \mu}(s) = {\tilde \mu}(1/2, s)
= - s \log\left[{1+{\varepsilon}^{2/s} \over 2}\right],$$we get the
following bound 
$$\begin{array}{ll}
E (R)\geq {\tilde \mu}({\tilde s}_R) - {\tilde s}_R R,\quad
& 0< R\leq {\tilde \mu}'(1);\\ E (R)\geq \tilde{C} - R, \quad &
{\tilde \mu}'(1)\leq R\leq {\mu}' (1);\\ E (R)\geq \mu(s_R ) - s_R R,
\quad & {\mu}'(1)\leq R <C,\end{array}$$
where ${\tilde s}_R, s_R$ are solutions of the equations
${\tilde \mu}' ({\tilde s}_R) = R,\quad {\mu}'(s_R) = R,$ and
$${\tilde C} = \mu (1) = {\tilde \mu} (1) = - \log \left({1+\varepsilon^2 \over 2}
\right),$$
$${\tilde \mu}' (1) = {\tilde \mu}(1) + {\varepsilon^2\log\varepsilon^2
\over 1+\varepsilon^2},$$
$${\mu}' (1) =
  - {(1-\varepsilon)^{2}\log \left({1-\varepsilon \over 2}\right) +
(1+\varepsilon)^{2}\log \left({1+\varepsilon \over 2}\right) 
\over 2(1+\varepsilon^{2})} \,,
$$ $$C = {\mu}'(0) = - \left[\left({1-\varepsilon \over 2}\right)
\log \left({1-\varepsilon \over 2}\right) + \left({1+\varepsilon \over
2}\right)
\log \left({1+\varepsilon \over 2}\right)\right].$$

The maximal amount of information $C_1$ obtainable with non-entangled (product)
measurements is attained for the uniform input probability distribution
($\pi = 1/2$) and the corresponding Bayes (maximum likelihood) decision rule given by the orthonormal basis in $\cal H$ oriented symmetrically with respect to vectors $|\psi_0\rangle , |\psi_1\rangle$ ( which in this particular case coincides with 
(\ref{iii1}))  \cite{xlev95}, \cite{xosa}.
It is equal to the capacity of classical binary symmetric channel with the error probability $(1- \sqrt{1-\varepsilon^2})/2$, that is
$$C_1 = {1 \over 2}\left[(1+\sqrt{1-\varepsilon^2})\log (1+\sqrt{1-\varepsilon^2})+
(1- \sqrt{1-\varepsilon^2})\log (1 - \sqrt{1-\varepsilon^2})\right] .$$
A comparison on this quantity with $\tilde {C}$ shows that $C_1 < \tilde{C}$ for $0<\varepsilon <1$ (although the difference between the two functions is quite small, see Fig. 3).
Since $C\geq\tilde{C}$, this implies strict superadditivity property $C_n > n
C_1 $ for the binary pure state channel with $0<\varepsilon <1$. However finding explicit quantum block codes realizing the  potential of strict superadditivity seems to be a difficult problem,
see \cite{xsasa}.
\vskip20pt \centerline{\bf \S 4. General signal states with finite entropy} %
\vskip10pt The general case is substantially more complicated already on the
level of quantum Bayes problem; in particular, so far no upper bound for the
average error probability is known, generalizing appropriately the
geometrically simple bound (\ref{iii3}). The proof given in \cite{xhol} (see
also \cite{xwest}) is based rather on a noncommutative generalization of the
idea of ``jointly typical'' sequences in classical theory \cite{xinf}. This
is realized by substituting in the average error probability (\ref{1-63})
the decision rule 
\begin{equation}
X_{w^{k}}=(\sum_{l=1}^{M}PP_{w^{l}}P)^{-1/2}PP_{w^{k}}P(%
\sum_{l=1}^{M}PP_{w^{l}}P)^{-1/2},  \label{gs1}
\end{equation}
where $P$ is the projector onto the typical subspace of 
\[
\bar{S}_{\pi }=\sum_{i\in A}\pi _{i}S_{i},
\]
and $P_{w^{k}}$ is a proper generalization of the typical projection for the
density operators $S_{w^{k}}$ . Namely, we choose $P_{w^{k}}$ to be the
spectral projection of $S_{w^{k}}$ corresponding to the eigenvalues $\lambda
_{J}$ in the interval $(\mbox{e}^{-n[{\bar{H}}_{\pi }(S_{(\cdot )})+\delta
]},\mbox{e}^{n[{\bar{H}}_{\pi }(S_{(\cdot )})-\delta ]})$. The essential
properties of $P_{w^{k}}$ are 
\begin{equation}
P_{w^{k}}\leq S_{w^{k}}\mbox{e}^{n[{\bar{H}}_{\pi }(S_{(\cdot )})+\delta ]},
\label{gs2}
\end{equation}
\begin{equation}
{\sf E}\mbox{Tr}S_{w^{k}}(I-P_{w^{k}})\leq \epsilon .  \label{gs3}
\end{equation}

The operator $(\sum_{l=1}^{M}PP_{w^{l}}P)^{-1/2}$ is to be
understood as generalized inverse of $(\sum_{l=1}^{M}PP_{w^{l}}P)^{1/2}$, equal to $0$ on the null subspace of that operator, which contains
range of the projector $I-P$.  Denoting $\hat{P}$ the projection onto the
range of $\sum_{l=1}^{M}PP_{w^{l}}P,$ we have
\begin{equation}  \label{P}
PP_{w^{l}}P \leq \hat{P} \leq P ,\quad l=1,\dots ,M.
\end{equation}

The proof given below is somewhat more direct
than that in \cite{xhol}, \cite{xwest}, making no use of eigenvectors and
spectral decompositions of the signal density operators $S_{w^{k}}.$

{\bf Theorem 3}. {\sl The capacity of a c-q channel $i\rightarrow S_i$
satisfying the condition (\ref{1-62}) is given by} 
\begin{equation}  \label{gs4}
C = {\bar C}\equiv \sup_{\pi} \Delta H( \pi).
\end{equation}

{\sl Proof}. We shall assume that the supremum is finite, otherwise the
modification is obvious. In view of the argument in \S II.3 we have only to
show that 
\begin{equation}  \label{R}
p(n, \mbox{e}^{nR}) \rightarrow 0 \qquad \mbox{for}\quad R < {\bar C}.
\end{equation}
To avoid cumbersome notations, we shall further enumerate words by the
variable $w$ omitting the index $k$.

By denoting ${A_w}=P_{w}P(\sum_{w'=1}^{M}PP_{w'}P)^{-1/2}$ and using the
inequality
\[
\left| \mbox{Tr}S_{w}{A_w}\right| ^{2}\leq \mbox{Tr}S_{w}{A_w}^{*}{A_w},
\]
we obtain
\[
{\bar{\lambda}}(W,X)\leq \frac{1}{M}\sum_{w=1}^{M}[1-\left| \mbox{Tr}%
S_{w}{A_w}\right| ^{2}]\leq \frac{2}{M}\sum_{w=1}^{M}[1-\mbox{Tr}S_{w}{A_w}],
\]
where $\mbox{Tr}S_{w}{A_w}=\mbox{Tr}PS_{w}P_{w}P(\sum_{w'=1}^{M}PP_{w'}P)^{-1/2}$ 
is real number between $0$ and $1$. Applying inequality
\[
-2x^{-1/2}\leq -3+x,\quad x > 0,
\]
which follows from (\ref{ner}), we obtain by (\ref{P})
\[
-2(\sum_{w'=1}^{M}PP_{w'}P)^{-1/2}\leq -3\hat{P}+\sum_{w'=1}^{M}PP_{w'}P
\leq -3P P_w P +\sum_{w'=1}^{M}PP_{w'}P.
\]
Hence
\[
{\bar{\lambda}}(W,X)\leq \frac{1}{M}\sum_{w=1}^{M}[2\mbox{Tr}S_{w}-3\mbox{Tr}%
S_{w}P_{w}PP_{w}P+\sum_{w^{\prime }=1}^{M}\mbox{Tr}S_{w}P_{w}PP_{w^{\prime }}P]
\]
\[
= \frac{1}{M}\sum_{w=1}^{M}[2\mbox{Tr}S_{w}(I-P_{w}PP_{w}P)+
\sum_{w^{\prime }:w^{\prime }\not{=}w}\mbox{Tr}S_{w}P_{w}PP_{w^{\prime }}P].
\]
Taking into account that 
\[
\mbox{Tr}S_{w}(I-P_{w}PP_{w}P)=\mbox{Tr}S_{w}(I-P_{w})PP_{w}P +
\mbox{Tr}S_{w}(I-P)P_{w} -
\mbox{Tr}S_{w}(I-P)P_{w}(I-P) + \]\[\mbox{Tr}S_{w}(I-P_{w})P +
\mbox{Tr}S_{w}(I-P) \leq %
2 [\mbox{Tr}S_{w}(I-P_{w})+\mbox{Tr}S_{w}(I-P)],
\]
we can write  
\begin{equation}
\inf_{X}{\bar{\lambda}}(W,X)\leq \frac{1}{M}\sum_{w=1}^{M}\{4\mbox{Tr}%
S_{w}(I-P)+4\mbox{Tr}S_{w}(I-P_{w})+\sum_{w^{\prime }:w^{\prime }\not{=}w}%
\mbox{Tr}PS_{w}PP_{w^{\prime }}\},  \label{gs10}
\end{equation}
which is our final basic bound.

We now again apply the Shannon's random coding scheme, assuming that the
words $w^1 ,..., w^M$ are chosen at random, independently and with the
probability distribution (\ref{iii5}) for each word. Then similarly to
(\ref{iii6}) ${\sf E}S_w = {\bar S}_{\pi}^{\otimes n}$ , where ${\bar
S}_{\pi} = \sum_{i\in A}\pi_i S_i $, and from (\ref{gs10}), by independence
of $S_w ,P_{w'}$, $${\sf E}\inf_{X}{\bar  \lambda}(W, X) \leq 4\mbox{Tr} {\bar
S}_{\pi}^{\otimes n} (I - P) + 4{\sf E}\mbox{Tr}S_w (I - P_w) + (M-1)\mbox{Tr}
{\bar S}_{\pi}^{\otimes n} P {\sf E}P_{w'} .$$ By the inequalities
(\ref{iii9}), (\ref{gs3}) expressing typicality of the projectors $P, P_w$,
and by the properties of trace, $${\sf E}\inf_{X}{\bar \lambda}(W, X)
\leq  8\epsilon + (M-1) \|  {\bar S}_{\pi}^{\otimes n} P\|\mbox{Tr} {\sf E}P_{w'},$$
for $n\geq n (\pi,\epsilon ,\delta )$.  By the property (\ref{iii8}) of $P$,
$$\| {\bar S}^{\otimes n} P\| \leq \mbox{e}^{-n[H({\bar S}_{\pi}) - \delta
]},$$ and by the property (\ref{gs2}) of $P_w$, $$\mbox{Tr} {\sf E}P_{w'} =
{\sf E}\mbox{Tr} P_{w'} \leq {\sf E}\mbox{Tr} S_{w'} \cdot \mbox{e}^{n[{\bar
H}(S_{(\cdot )}) + \delta ]} = \mbox{e}^{n[{\bar H}(S_{(\cdot )}) + \delta
]}.$$ Thus $${\sf E}\inf_{X}{\bar \lambda}(W, X) \leq 8\epsilon + (M-1)
\mbox{e}^{-n[H({\bar S}_{\pi}) - {\bar H}(S_{(\cdot )}) - 2\delta ]}. $$

Let us choose the distribution $\pi = \pi^0$ such that $\Delta H(\pi^0 )\geq
{\bar C} - \delta$. Then \begin{equation}\label{gs11} p(n, M)\leq 8\epsilon +
(M-1)
\mbox{e}^{-n[{\bar C} - 3\delta ]} \end{equation} for $n\geq n(\pi^0 , \epsilon
,\delta )$.  Thus $p(n, \mbox{e}^{n[{\bar C} - 4\delta ]})\rightarrow 0$ as
$n
\rightarrow
\infty$, whence (\ref{R}) follows.$\Box$

For quasiclassical channel where the signal states are given by commuting density
operators $S_i$ one can use the classical bound
of Theorem 5.6.1 \cite{xgal} with
transition probabilities $S(\omega |i)$, where $S(\omega |i)$ are the
eigenvalues of $S_i$.  In terms of the
density operators it takes the form
\begin{equation}\label{gs12}      
{\sf E}\inf_{X}\bar{\lambda}(W, X) \leq \inf_{0\leq s\leq 1}
 (M-1)^{s}\left(\mbox{Tr}\left[ \sum_{i\in A}
\pi_i S_i^{1 \over 1+s}\right] ^{1+s}\right) ^n.\end{equation}
The righthand side of (\ref{gs12}) is meaningful for arbitrary
density operators, which gives a hope that this estimate, with some modification, could be
generalized to the noncommutative case (note that for pure states $S_i$
Theorem 1 gives twice the expression (\ref{gs12})).
This would not only give a
different proof of Theorem 3, but also a lower bound for the quantum
reliability function in the case of general signal states, possibly with
infinite entropy.
\vskip20pt
\centerline{\sc IV. Quantum channels with constrained inputs}\vskip10pt\centerline{\bf \S 1.
The case of discrete alphabet}
\vskip10pt
Importance of quantum channels with constrained inputs was clear from the
beginning of quantum communication; the question ``How little energy is
needed to send a bit?'' is formulated more precisely as calculation of the
capacity of quantum channel with constrained input energy (see
\cite{xgordon}, \cite{xle}, \cite{xbowen}, \cite{xcaves} for more physical discussion on that point).

We first consider the case of discrete alphabet $A=\{i\}$. Let $f(i)$ be a
nonnegative function defined on the input alphabet. We shall consider the
class ${\cal P}_1$ of input distributions $\pi$ satisfying the condition
\begin{equation} \label{1a-1}
\sum_i f(i) \pi(i) \leq E ,
\end{equation}
where $E$ is a real number.

We put the additive constraint onto the input words $w = (i_1 ,...,i_n )$ by
asking
\begin{equation} \label{1a-2}
f(i_1)+\ldots + f(i_n)
\leq nE ,\end{equation}
and denote by ${\cal P}_n$ the class of probability distributions satisfying
the corresponding condition
\begin{equation} \label{1a-3}
\sum_{i_1, \ldots, i_n} [f(i_1)+\ldots + f(i_n)] \pi (i_1,\ldots,i_n)
\leq nE .\end{equation}

Now the quantities $C, {\bar C}$ can be defined as in \S II.2 with the
modification that the suprema with respect to $\pi$ are taken over ${\cal
P}_n$, that is $$C = \lim_{n \to \infty} C_n /n ,\qquad {\bar C} = \lim_{n
\to \infty} {\bar C}_n /n,$$
where $$C_n = \sup_{\pi\in{\cal P}_n} \sup_{X} I_n (\pi , X),\qquad {\bar
C}_n = \sup_{\pi\in{\cal P}_n} \Delta H_n(\pi ) .$$ and $ I_n (\pi , X),
\Delta H_n(\pi )$ are the analogues of the mutual information (\ref{1-81}) and the entropy bound (\ref{1-8}) for the product c-q channel in ${\cal H}^{\otimes n}$.

Let us remark that just as it was in the case of unconstrained inputs, the
sequence ${\bar C}_n$ is additive and hence
\begin{equation} \label{1a-9}
{\bar C} = \sup_{\pi\in{\cal P}_1} \Delta H(\pi ) .\end{equation} Indeed, it
is sufficient to check that \begin{equation}\label{leq} {\bar C}_n \leq n
{\bar C}_1.  \end{equation} By the subadditivity of quantum entropy with
respect to tensor products, $$\Delta H_n (\pi )\leq
\sum_{k=1}^n \Delta H (\pi^{(k)} ),$$ where $\pi^{(k)}$ is the $k$-th
marginal distribution of $\pi$ on $A$. Also $$
\sum_{k=1}^n \Delta H (\pi^{(k)} )
\leq n \Delta H ({\bar \pi}),$$ where ${\bar \pi} = {1 \over
n}\sum_{k=1}^n \pi^{(k)}$, since $\Delta H (\pi )$ is concave function of
$\pi$ \cite{xhol73}.The inequality (\ref{1a-3}) can be rewritten as $$ {1
\over n}\sum_{k=1}^n \sum_{i_k}f(i_k ) \pi^{(k)}(i_k) \leq E ,$$
which implies that ${\bar \pi}\in {\cal P}_1$ if ${\pi}\in {\cal P}_n$ .
Taking supremum with respect to $\pi\in{\cal P}_n$ proves (\ref{leq}).

{\bf Theorem 4.} {\sl The capacity of a c-q channel $i\rightarrow S_i$
satisfying the condition (\ref{1-62}) with the input constraint (\ref{1a-2})
is equal to (\ref{1a-9})}.

{\sl Proof}. We have to show that if $C < \infty$ and $R > C$ , then $ p(n,
\mbox{e}^{nR})\not\rightarrow 0$, and if the condition (\ref{1-62}) holds and
$R < \bar{C}$ , then $p(n, \mbox{e}^{nR}) \rightarrow 0$ .

The proof of the first statement (the converse coding theorem) is based on
the following modification of the inequality (\ref{1-10})
\begin{equation}\label{1a-10}
\log M \cdot(1 - p(n, M))\leq \sup_{\pi\in{\cal P}_n} \sup_X I_n (\pi , X) + 1,
\end{equation}
Let again as in the proof of Theorem 1 $\cal J$ be the classical random
variable describing the output of the product channel under the decision rule
$X$ if the words in the code $(W, X)$ are taken with the input distribution
$\pi_M$ assigning equal probability $1/M$ to each word. Consider Fano
inequality.  Since the words in the code satisfy (\ref{1a-2}), we have
$\pi_M\in{\cal P}_n$, and (\ref{1a-10}) follows by taking supremum with
respect to $(W, X). $ Substituting $M=
\mbox{e}^{nR}$ gives the first statement of the Theorem.

In the classical information theory direct coding theorems for channels with
additive constraints are proved by using random coding with probability
distribution (\ref{iii5}) modified with a factor concentrated on words, for
which the constraint holds close to the equality \cite{xgal}, Sec. 7.3.  The
same tool can be applied to quantum channels \cite{xhol98}. Let $\pi$ be a
distribution satisfying (\ref{1a-1}), and let {\sf P} be a distribution on
the set of $M$ words, under which the words are independent and have the
probability distribution (\ref{iii5}).  Let $\nu_n = {\sf P} ({1 \over
n}\sum_{k=1}^n f(i_k) \leq E)$ and define the modified distribution under
which the words are still independent but
\begin{equation}\label{1a-11}     {\tilde {\sf P}} (w = (i_1,\ldots,i_n)) =
\left\{\begin{array}{ll}
\nu_n^{-1}  \pi_{i_1}\cdot\ldots\cdot\pi_{i_n}, & \mbox{if}\,
\sum_{k=1}^n f(i_k)\leq nE,\\0, &
\mbox{otherwise.}\end{array}
\right. \end{equation} Let us remark that since $\pi\in {\cal P}_1$, then ${\sf E}f
\leq E$ (where {\sf E} (${\tilde {\sf
E}}$) is the expectation corresponding to {\sf P} (${\tilde {\sf P}}$)) and
hence by the central limit theorem $$\lim_{n\rightarrow\infty}\nu_n\geq 1/2
.$$ Therefore
${\tilde {\sf E}} \xi \leq 2^m {\sf E} \xi$ for any nonnegative random
variable $\xi$ depending on $m$ words.

For the error probability (\ref{1-63}) we have the basic upper bound
(\ref{gs10}).  Take the expectation of this bound with respect to $ {\tilde
{\sf P}}$.  Since every summand in the right hand side of (\ref{gs10})
depends no more than on two different words, we have $${\tilde {\sf E}}
\inf_{X}{\bar \lambda}(W, X)\leq 4  {\sf E} \inf_{X}{\bar \lambda}(W, X),$$
and the expectation with respect to {\sf P} can be made arbitrarily small
provided $M=\mbox{e}^{nR}, n\rightarrow\infty,$ with $R < C - 3\delta$. Thus
${\tilde {\sf E}}{\bar \lambda}$ also can be made arbitrarily small under the
same circumstances.  Since the distribution $\tilde P$ is concentrated on
words satisfying (\ref{1a-2}), we can choose a code for which $\bar \lambda
(W, X)$ can be made arbitrarily small.  $\Box$

\vskip20pt
\centerline{\bf \S 2. The case of continuous alphabet}
\vskip10pt

In this section we take as the input alphabet $A$ arbitrary Borel subset in a
finite-dimensional Euclidean space $\cal E$.

We assume that a nonnegative Borel function $f$ on $\cal E$ is fixed and
consider the set ${\cal P}_1$ of probability measures $\pi$ on $A$ satisfying
\begin{equation} \label{2-0}
\int_{A} f(x) \pi (dx) \leq E .
\end{equation}
We impose the additive constraint onto transmitted words $w = (x_1,...,x_n)$
by requiring
\begin{equation}\label{2-1}
f (x_1) + ... + f(x_n)\leq E.  \end{equation} Like in the classical case, we
discretize the channel by taking apriori distributions with discrete supports

\begin{equation} \label{2-2}
\pi(dx) = \sum_{i}\pi_i \delta_{x_i} (dx),\end{equation} where $\{
x_i\}\subset A$ is arbitrary countable collection of points and $$
\delta_x (B) =\left\{\begin{array}{ll}1, & \mbox{if}\quad x\in B,\\
0, & \mbox{if} \quad x \not\in B,\end{array}\right.$$ 
For $\pi$ of the form (\ref{2-2}) the condition
(\ref{2-0}) takes the form (\ref{1a-1}).  Denoting ${\cal P}_1'$ the class of
all such probability distributions, we can directly extend the argument of
Theorem 4, with the capacity given by
\begin{equation}\label{2-3}C = \bar{C}\equiv\sup_{\pi\in{\cal P}_1'} \Delta H(\pi ) .
\end{equation}

We now assume that the channel is given by {\sl weakly continuous} mapping
$x\rightarrow S_x$ from the input alphabet $A$ to the set of density
operators in $\cal H$. (The weak continuity means continuity of all matrix
elements $\langle\psi |\, S_x\, \phi \rangle; \psi,\phi \in \cal H$).  For
arbitrary $\pi $ consider the quantity
\begin{equation} \label{2-4}
\Delta H (\pi ) =  \int_A H (S_x ; {\bar S}_{\pi}) \pi (dx) ,
\end{equation}
where\begin{equation} \label{2-5} {\bar S}_{\pi} = \int_A S_x \pi (dx).
\end{equation}
Because of the weak continuity of the function $S_x$ the integral is well
defined and represents a density operator in $\cal H$. Moreover, the
nonnegative function $H(S_x ; {\bar S}_{\pi})$ is lower semicontinuous (see
\cite{xwehrl}),
and hence the integral in (\ref{2-4}) is also well defined .

We also introduce the analog of the condition (\ref{1-62}):
\begin{equation}\label{2-5a}
\sup_{x \in A} H(S_x) < \infty .      \end{equation}
Under this condition the representation (\ref{1-8}) holds with $${\bar H}
(S_{(\cdot )} ) = \int_A H (S_x ) \pi (dx) .$$

{\bf Proposition 2}. {\sl Under the assumption that the mapping $x\rightarrow
S_x$ is weakly continuous, the function $f$ is lower semicontinuous, and the
condition (\ref{2-5a}) holds,
\begin{equation}
\label{2-6} {\bar C} = \sup_{\pi \in {\cal P}_1} \Delta H (\pi ) . \end{equation}

Proof. } In view of (\ref{2-3}) we have only to show that $$ \sup_{\pi \in
{\cal P}_1'} \Delta H (\pi ) \geq \sup_{\pi \in {\cal P}_1} \Delta H (\pi
).$$ It is sufficient to construct, for arbitrary $\pi\in {\cal P}_1$, a
sequence of ${\pi}^{(l)} \in {\cal P}_1'$ such that
\begin{equation} \label{2-7}
\lim\inf_{l\rightarrow \infty} \Delta H (\pi^{(l)}) \geq \Delta H (\pi
) .\end{equation} To this end for any $l =1, \ldots$ we consider the division
of $A$ into disjoint subsets
\begin{equation} \label{2-9}
B_k^{(l)} = \{ x: k/l\leq H( S_x )<(k+1)/l \},\quad k=\ldots, -1, 0, 1,
\ldots\, .\end{equation} By making, if necessary, a
finer subdivision, we can always assume that diameters of all sets
$B_k^{(l)}$ are bounded from above by $\epsilon_l$, where
$\epsilon_l\rightarrow 0$ as $l\rightarrow\infty$.  Let $x_k^{(l)}$ be a
point at which $f(x)$ achieves its minimum on the closure ${\bar B_k^{(l)}}$
of $B_k^{(l)}$, and define
\begin{equation} \label{2-10}
\pi^{(l)} (dx) = \sum_k \pi (B_k^{(l)}) \delta_{x_k^{(l)}} (dx) ,
\end{equation}
where $\pi$ is a fixed distribution from ${\cal P}_1$.  Then $$\int_A f(x)
\pi^{(l)}(dx) \leq \int_A f(x) \pi (dx),$$ hence
$\pi^{(l)}\in {\cal P}_1'$.

By construction (\ref{2-9}), (\ref{2-10}) we have
\begin{equation}\label{2-11}      \left|\int_A H( S_x  ) \pi^{(l)} (dx) - \int_A H( S_x  ) \pi (dx)\right| \leq
1/l .\end{equation} Let us show that \begin{equation} \label{2-12}
\lim\inf_{l\to\infty} H\left(\int_A  S_x  \pi^{(l)} (dx)\right) \geq
H\left(\int_A S_x \pi (dx)\right) .\end{equation}

To this end we remark that due to the weak continuity and uniform boundedness
of the function $S_x$, the operators $\int_A S_x
\pi^{(l)} (dx)$ weakly converge to the  operator $\int_A  S_x
\pi (dx)$. Indeed, let $B_c$ be the ball of radius $c$ in ${\cal E}$.
Then $$\left| \langle\phi | \int_A S_x
\pi^{(l)} (dx) \psi\rangle - \langle\phi | \int_A  S_x
\pi (dx) \psi\rangle \right| $$ $$\leq \sum_k \int_{B_k^{(l)}\cap B_c}
|\langle\phi | S_{x_k^{(l)}}\psi\rangle - \langle\phi | S_x \psi\rangle|
\pi (dx) + 2\|\phi\| \|\psi\| \pi (A\setminus B_c) .$$ By choosing first $c$
large enough to make the second term small, we can make the first term small
for all large enough $l$ since $\langle\phi | S_x \psi\rangle$ is uniformly
continuous on $A\cap B_c$ and the diameters of $B_k^{(l)}$ uniformly tend to
zero. The relation (\ref{2-12}) then follows from the lower semicontinuity of
the quantum entropy.  Relations (\ref{2-11}), (\ref{2-12}) imply (\ref{2-7}).
$\Box$
\vskip20pt
\centerline{\bf \S 3. The upper bounds for error probability}
\vskip10pt

A much more detailed information concerning the rate of convergence of
the error probability can be obtained for pure state channels, by
modifying the estimates from \S III.2 to channels with infinite
alphabets and constrained inputs following the method of \cite{xgal},
Ch. 7.  We start with the case of discrete alphabet $A$.

Let $S_i = |\psi_i\rangle\langle\psi_i|$ be the pure signal states of the channel,
and let $\pi$ be an apriori distribution satisfying the condition
(\ref{1a-1}). Then the
following {\sl random coding bound} holds for the error probability $p(n,
M)$ where $M = \mbox{e}^{nR}$ with $R < C$:
\begin{equation} \label{40}
p(n, \mbox{e}^{nR}) \leq 2 \left({\mbox{e}^{p\delta} \over \nu_{n,
\delta }}\right)^2 \mbox{exp}\{-n[{\mu (\pi, s, p) - s R]}\},
\end{equation}
where \begin{equation} \label{41}
\mu (\pi, s, p) = - \log \mbox{Tr}\left\{\sum_{i\in A} \pi_i \mbox{e}^{p[f(i)
- E]} S_i\right\}^{1+s}, \end{equation}
and $0\leq s\leq 1, 0\leq p, 0 < \delta $ are arbitrary parameters. 
The quantity $$\nu_{n, \delta } =  {\sf P}(En - \delta \leq 
\sum_{k=1}^n f(i_k)\leq nE )$$ satisfies
$\lim_{n\to \infty}\sqrt{n}\nu_{n,\delta } > 0$, thus adding only
$o(n)$ to the exponential in (\ref{40}).

The bound (\ref{40}) is obtained in the same way as Theorem 2, that is by evaluating the expectation of the average error
probability (\ref{V1}) using random, independently chosen codewords, but
with the modified codeword distribution
\begin{equation} \label{42}
{\tilde {\sf P}}_{\delta} (u = (i_1,\ldots,i_n)) = 
\left\{\begin{array}{ll}\nu_{n,\delta }^{-1}\,  
\pi_{i_1}\cdot\ldots\cdot\pi_{i_n}, & \mbox{if}\, nE - \delta \leq 
\sum_{k=1}^n f(i_k)\leq nE,\\
0, & \mbox{otherwise.}\end{array} \right.\end{equation}
By using independence of the words, we can repeat the first part 
of the proof of Theorem 2 to show that
\[
-2\,{\tilde {\sf E}}_{\delta}G^{1/2}\leq -2{\tilde {\sf E}}_{\delta}G+\left\{ 
\begin{array}{l}
2\,{\tilde {\sf E}}_{\delta}G \\ 
{\tilde {\sf E}}_{\delta}(G^{2}-G)\end{array}
\right. =
 -2M{\tilde {\sf E}}_{\delta}S_w+\left\{ 
\begin{array}{l}
2M{\tilde {\sf E}}_{\delta}S_w \\ M(M-1)
({\tilde {\sf E}}_{\delta}S_w )^{2}
\end{array}
\right. .
\]
Now for any $p\geq 0$,
$${\tilde {\sf E}}_{\delta}\,S_w \leq \left({\mbox{e}^{p\delta} \over
\nu_{n,\delta } }\right)\, {\sf E}\mbox{exp}\{p\sum_{k=1}^n [f(i_k) -
E]\} \,S_w = \left({\mbox{e}^{p\delta} \over
\nu_{n,\delta } }\right)\, \left(\sum_{i\in A}\mbox{e}^{p[f(i) -
E]} \pi_i S_i\right)^{\otimes n}\equiv \tilde{\Sigma} .$$
Then we obtain for $0\leq s\leq 1$
$$ \frac{2}{M}(M-{\tilde {\sf E}}_{\delta}G^{1/2})\leq
2 (M-1)^s {\rm Tr} \tilde{\Sigma}^{1+s}\leq 2 \left({\mbox{e}^{p\delta} \over
\nu_{n,\delta } }\right)^2\, \left(\mbox{Tr}\left\{\sum_{i\in A}\mbox{e}^{p[f(i) -
E]} \pi_i S_i\right\}^{1+s}\right)^{ n} ,$$ whence
 the bound (\ref{40}) follows.

In the same way, the {\sl expurgated bound} from \S III.2 can be
modified to obtain 
\begin{equation} \label{43}
p(n, \mbox{e}^{nR}) \leq  \mbox{exp}\{-n[{\tilde \mu} (\pi, s, p) - s (R + {2
\over n}\log{2\mbox{e}^{p\delta} \over \nu_{n,\delta}})]\},
\end{equation}
where \begin{equation} \label{44}
{\tilde \mu} (\pi, s, p) = - s \log \sum_{i,k\in A}\pi_i\pi_k\mbox{e}^{p[f(i)
+ f(k) - 2E]} \,|\langle\psi_i |\psi_k\rangle|^{2/s}.\end{equation}

These bounds can be extended to pure state channels with continuous
alphabets by using technique of \S 2 to obtain (\ref{40}), (\ref{43}) with
\begin{equation} \label{45}
\mu (\pi, s, p) = - \log \mbox{Tr}\left\{\int_A \mbox{e}^{p[f(x)
- E]} S_x \pi (dx) \right\}^{1+s}, \end{equation}
\begin{equation} \label{46}
{\tilde \mu} (\pi, s, p) = - s \log \int_A\int_A \mbox{e}^{p[f(x)
+ f(y) - 2E]} |\langle\psi_x |\psi_y\rangle|^{2/s}\pi (dx) \pi (dy).\end{equation}

Introducing the {\sl reliability function} (\ref{rel}), we get the lower bound for $E(R)$:
$$E(R)\geq \max \{ E_r (R), E_{ex} (R)\},$$
where
\begin{equation} \label{47}
E_r(R) = \sup_{ 0\leq s \leq 1} 
(\sup_{0\leq p}\sup_{\pi\in{\cal P}_1}\mu (\pi, s, p) -sR), \end{equation}
\begin{equation} \label{48}
E_{ex}(R) = \sup_{ 1\leq s} 
(\sup_{0\leq p}\sup_{\pi\in{\cal P}_1}{\tilde \mu} (\pi, s, p) -sR). 
\end{equation} An example where the
maximization at least partially can be performed analytically will be
considered in \S V.2.
\vskip20pt
\centerline{\sc V. Quantum Gaussian channels}\vskip10pt\centerline{\bf \S 1. Gaussian memoryless channel in one mode.}\vskip10pt
For a simple introduction to Gaussian states see e. g. \cite{xprob}, Ch. V.
To make presentation self-contained, we include proofs of some well known
results (such as Lemma 1 below).

Let $A$ be the complex plane {\bf C}, and let for every $\alpha \in {\bf C}$
the density operator $S_\alpha $ describe the thermal state of harmonic
oscillator with the signal amplitude $\alpha $ and the mean number of the
noise quanta $N$, i. e.
\begin{equation} \label{3-1}
S_\alpha ={1 \over \pi N} \int\mbox{exp}\left( - {|z-\alpha |^2 \over
N}\right) |z\rangle\langle z| d^2 z ,\end{equation} where $|z\rangle$ are the
coherent state vectors. By introducing the creation - annihilation operators
$a^{\dagger}, a$, we have
\begin{equation}\label{3-2}
\mbox{Tr}S_\alpha a =\alpha ,\qquad
\mbox{Tr}S_{\alpha} a^{\dagger}a = N + |\alpha |^2 .
\end{equation}
We remind that $$S_{\alpha} = V(\alpha ) S_0 V(\alpha )^*,$$ where $$V(\alpha
) = \mbox{exp}(\alpha a^{\dagger} - {\bar
\alpha}a)$$ are the unitary displacement operators, and the operator $S_0$ has
the spectral representation:
\begin{equation}\label{3-3}
S_0 ={1 \over N+1} \sum_{n=0}^{\infty}\left({N \over N+1}\right)^n
|n\rangle\langle n|,
\end{equation}
where $|n\rangle$ are the eigenvectors of the number operator $a^{\dagger}a$,
corresponding to eigenvalues $n=0,1,\ldots $.  Hence the states (\ref{3-1})
all have the same entropy
\begin{equation} \label{3-4}
H ( S_\alpha ) = (N+1)\log (N+1) - N\log N \equiv g(N) ,\end{equation} where
$g(x)$ is continuous concave monotonously increasing function of $x\geq 0$.
It is well known that the states $ S_\alpha $ have maximal entropy among all
states satisfying (\ref{3-2}). This follows from

{\bf Lemma 1}. {\sl The operator (\ref{3-3}) has maximal entropy among all
density operators $S$ satisfying}
\begin{equation}\label{3-2a}
\mbox{Tr}S a^{\dagger}a \leq N .
\end{equation}

{\sl Proof}. Denote $S'$ the density operator which is diagonal in the basis
$\{|n\rangle\}$ with the elements $s_n = \langle n |S| n\rangle$ on the
diagonal.  This operator satisfies (\ref{3-2a}), and $$H(S') - H(S) = H(S;
S')\geq 0.$$ Therefore the maximum is achieved on the diagonal operators. One
must maximize $H(S) = - \sum_n s_n\ln s_n$ under the conditions $s_n\geq 0,
\sum_n s_n = 1$, and
(\ref{3-2a}) which becomes $\sum_n n s_n\leq N$, and the solution (\ref{3-3})
follows by application of the Lagrange method.$\Box$

Operator (\ref{3-3}) satisfies the conditions (\ref{3-2}) (with
$\alpha= 0$) therefore it also has the maximal entropy among such density
operators.

Let us consider the channel $\alpha\rightarrow S_{\alpha}$ which is quantum
analog of memoryless channel with additive Gaussian noise (see
\cite{xgordon}, \cite{xhel},
\cite{xhol77}).
The condition (\ref{2-5a}) is trivially satisfied, and for any input
distribution $\pi (d^2 \alpha )$ $$\Delta H (\pi ) = H({\bar S}_{\pi}) - H (
S_0) ,$$ where $${\bar S}_{\pi} =
\int S_\alpha\,  \pi (d^2 \alpha ) .$$
By choosing $f(\alpha ) = |\alpha |^{2}$, we impose the additive constraint
of the type (\ref{2-1}). Thus ${\cal P}_1$ is defined as
\begin{equation} \label{3-5}\int |\alpha |^2 \,\pi (d^2 \alpha ) \leq
E ,\end{equation} where $\pi (d^2 \alpha )$ is an input distribution.  In
fact, $E$ is the ``mean number of quanta'' in the signal, which is
proportional to energy for one mode.  The constraint (\ref{3-5}) by virtue of
(\ref{3-2}) implies \begin{equation} \label{3-6}
\mbox{Tr} \, {\bar S}_{\pi}\, a^{\dagger} a \leq N+E .
\end{equation}
According to Lemma 1 the maximal entropy
\begin{equation} \label{3-7}
H( {\bar S}_{\pi}) = g(N+E)
\end{equation}
is attained by Gaussian density operator
\begin{equation} \label{3-8}
{\bar S}_{\pi} ={1 \over \pi (N+E)} \int\mbox{exp}\left( - {|z|^2 \over
(N+E)}\right) |z\rangle\langle z| d^2 z ,\end{equation} corresponding to the
optimal distribution
\begin{equation} \label{3-9}\pi
(d^2 \alpha ) = {1 \over \pi E}\, \mbox{exp}\left( - {|\alpha |^2 \over
E}\right) d^2 \alpha .\end{equation} By Proposition 2 the capacity of the
memoryless Gaussian channel is given by $$C = {\bar C}\equiv g(N+E) - g(N) $$

$$=\log \left(1 + {E \over N+1}\right) + (N+E)\log \left(1 + {1 \over
N+E}\right) - N \log \left(1 + {1 \over N}\right).$$

This quantity was anticipated in \cite{xgordon} (relation (4.28)) as an upper
bound for the information transmitted by the quantum Gaussian channel. On the
other hand, for a long time this quantity was also known as the capacity of
the ``narrow band photon channel'' \cite{xgordon62}, \cite{xle},
\cite{xcaves}. Our argument based on
Theorem 4 and Proposition 2 gives for the first time the proof of the
asymptotic equivalence, in the sense of information capacity, of the
memoryless Gaussian channel with the energy constraint to this quasiclassical
channel. To make the point clear, we give below a simplified one-mode
description of the photon channel.

Consider the discrete family of states
\begin{equation} \label{3-10}
S_m = P( m ) S_0 P( m )^*,\qquad m = 0, 1,\dots , \end{equation} where $P(m)$
is energy shift operator satisfying $ P(m)|n\rangle =|n+m\rangle.$ Notice
that $P( m ) = P^m$, where $P$ is isometric operator adjoint to the
quantum-mechanical ``phase operator'' \cite{xprob}. The states $S_m$ all have
the same entropy (\ref{3-4}) as the states $S_{\alpha}$, and the mean number
of quanta
\begin{equation} \label{3-11}
\mbox{tr} S_m \, a^{\dagger}a = N + m .
\end{equation}
Moreover, all states (\ref{3-10}) are diagonal in the number representation,
so the channel is quasiclassical.

Imposing the constraint
\begin{equation} \label{3-12}
\sum_{m=0}^{\infty} m \, \pi_m \leq E ,\end{equation}
where $\pi_m$ is an input distribution, and introducing the density operator
$${\bar S}_{\pi}' = \sum_{m=0}^{\infty} \pi_m S_m, $$ by virtue of
(\ref{3-11}), we obtain the same constraint (\ref{3-6}) for the new operator
${\bar S}_{\pi}'$. The maximal entropy (\ref{3-7}) is again attained by the
operator (\ref{3-8}), which has the spectral representation
\begin{equation}\label{3-13}
{\bar S}_{\pi} = {1 \over N+E+1} \sum_{n=0}^{\infty}\left({N+E \over
N+E+1}\right)^n |n\rangle\langle n|.
\end{equation}
It corresponds to the optimal signal distribution \cite{xle} $$\pi_m = {N
\over N+E} \delta_{m0} + {E \over N+E}\left[{1 \over
N+E+1} \left({N+E \over N+E+1}\right)^m\right] .$$

There is notable difference between the case of pure state channel as opposed
to the general case. For a pure state case (where $N = 0$), one can formulate
a broader problem of finding a maximum capacity channel $x\rightarrow S_x$
with arbitrary alphabet $\{x\}$ and a signal distribution $ \pi (dx)$
satisfying the output constraint $$\mbox{Tr}{\bar S}_{\pi}\, a^{\dagger} a
\leq E .$$
This was done in \cite{xyuen} where it was shown that the noiseless photon
channel provides a solution to this problem. In view of the result of
\cite{xjozsa}, any other pure state channel satisfying
$$\int S_x \pi (dx) = {1 \over E+1} \sum_{n=0}^{\infty}\left({E \over
E+1}\right)^n |n\rangle\langle n| $$ gives, asymptotically, a solution to the
same problem.However, in the general case imposing the output constraint
(\ref{3-6}) instead of the input constraints (\ref{3-5}) or (\ref{3-12})
looks rather artificial; the equivalence of these constraints for apparently
different channels seems to be a very special feature of the quantum Gaussian
density operators.

\vskip20pt
\centerline{\bf \S 2. The reliability function of  Gaussian
pure state channel}
\vskip10pt

We are going to apply results of  \S IV.3 to the
Gaussian pure state channel $\alpha \rightarrow  S_{\alpha} = 
|\alpha\rangle\langle\alpha|$ with the constraint  (\ref{3-5}). By  taking the optimal
apriori distribution (\ref{3-9}) we can calculate explicitly the
functions (\ref{45}), (\ref{46}).  

Namely, to calculate (\ref{45}), we remark that$$
\int\mbox{e}^{p(|z|^2 - E)} S_z \pi(d^2 z) = {\mbox{e}^{-pE} \over 1-pE}\,
{1 \over \pi E'}\int \mbox{e}^{-{|z|^2 \over E'}} |z\rangle\langle z| d^2 z$$ $$ =
 {\mbox{e}^{-pE} \over 1-pE}\,
{1 \over E'+1}\sum_{n=0}^{\infty}\left({E' \over E'+1}\right)^n |n\rangle\langle n|,$$
where $E' = E/(1-pE)$, provided $p < E^{-1}$, 
and the trace of the $(1+s)$-th power of this operator is easily
calculated to yield
\begin{equation} \label{50}
\mu (\pi, s, p) = (1+s)pE + \log [(1+E - pE)^{1+s} -
E^{1+s}].\end{equation}
 
By taking into account that $$|\langle z|w\rangle|^2 = \mbox{e}^{-|z-w|^2},$$
(see, e. g. \cite{xhel}), we can calculate the integral in (\ref{46}) as
$${\mbox{e}^{-2pE} \over (\pi E)^2}\int\int \exp\{ - [(E^{-1}+s^{-1}-p)|z|^2 
+ (E^{-1}+s^{-1}-p)|w|^2 -2s^{-1}\mbox{Re}{\bar z}w ]\}$$ 
$$ = {\mbox{e}^{-2pE} \over 1 + p^2E^2 -2pE-2pE^2/s + 2E/s},$$
for $p < E^{-1}$, whence 
\begin{equation} \label{51}
{\tilde \mu} (\pi, s, p) = s\{2pE + \log [1+p^2E^2 -2pE +
2E(1-pE)/s]\}.
\end{equation}

Trying to maximize $\mu (\pi, s, p)$ with respect to $p$ we obtain the
equation \begin{equation} \label{52}
(1 + E - pE)^s (1 - p) = E^s ,\end{equation}
which can be solved explicitly only for $s =0,1$. Thus, contrary to
the classical case \cite{xgal}, the maximum in (\ref{47}) in general
can be found only numerically. For $s=0$ we have $p=0$ and
$$C = \frac{\partial}{\partial s}\mu (\pi, 0, 0) = (E+1)\log (E+1) -
E\log E.$$
For $s=1$ equation (\ref{52}) has the unique solution 
$p(1, E) = 1 + 1/E - q(E)/E < E^{-1}$, where 
$$q(E) = {1 + \sqrt{4E^2 + 1} \over 2}.$$
For future use we find the important quantities
$$
\mu (\pi, 1, p(1, E)) = 2(E + 1 - q(E)) + \log q(E);$$
\begin{equation} \label{53}
\frac{\partial}{\partial s}\mu (\pi, 1, p(1, E)) = E + 1 - q(E) + 
{q(E)^2\log q(E) -E^2 \log E \over q(E)^2 - E^2}. \end{equation}

The optimization of the expurgated bound can be
performed analytically. Taking partial derivative with respect to
$p$ we obtain the equation
$$p^2 - 2p\left({1 \over s} + {1 \over 2E}\right) + {1 \over sE} = 0,$$ 
the solution of which, satisfying $p < E^{-1}$, is
$$p(s, E) = s^{-1} + E^{-1} - E^{-1}q(E/s).$$ 
Substituting this in (\ref{48}), we obtain the following expression,
which is to be maximized with respect to $s\geq 1$:
$${\tilde \mu}(\pi, s, p(s, E)) - sR = 2(E + s - s q(E/s)) + s\log q(E/s)
- sR.$$ Taking derivative with respect to $s$, we obtain the equation
$$q(E/s) = \mbox{e}^R,$$ the solution of which is 
\begin{equation} \label{54} s = {E \over
\sqrt{\mbox{e}^{2R} - \mbox{e}^{R}}}. 
\end{equation}
If this is less than 1, which
is equivalent to $$R < \log q(E) = \frac{\partial}{\partial s}{\tilde
\mu} (\pi, 1, p(1,E)),$$ then   the maximum is achieved for the value of
$s$ given by (\ref{54}) and is equal to
$$2E (1 - \sqrt{1 - \mbox{e}^{-R}}) = E_{ex} (R) > E_r (R),$$
(which up to a factor coincides with the expurgated bound for
classical Gaussian channel). In the range 
$$ \frac{\partial}{\partial s}{\tilde
\mu} (\pi, 1, p(1,E)) \leq R \leq \frac{\partial}{\partial s}
\mu (\pi, 1, p(1,E)), $$where the optimizing $s$ is equal to 1, 
we have the linear bound$$E_{ex} (R) = E_r (R) = \mu (\pi, 1, p(1,E)) - R,$$
with the quantities $\frac{\partial}{\partial s}
\mu (\pi, 1, p(1,E)), \mu (\pi, 1, p(1,E))$ defined by
(\ref{53}). Finally, in the range $$\frac{\partial}{\partial s}
\mu (\pi, 1, p(1,E)) < R < C$$ we have $E_{ex} (R) < E_r (R)$ with $E_r
(R)$ given implicitly by (\ref{47}).

On the other hand, for the pure state photon channel the analysis of
the error probability is trivial: since this is quasiclassical noiseless
channel, the error probability is zero for $R < C$. Thus, although the
two channels are asymptotically equivalent in the sense of capacity,
their finer asymptotic properties are apparently essentially different.
\vskip20pt
\centerline{\bf \S 3. The case of many modes}
\vskip10pt
Consider now a finite collection of harmonic oscillators with frequencies
$\omega_j$, described by creation-annihilation operators $a_j^{\dagger}, a_j
; j= 1, ...$~.  Let $\alpha_j\in {\bf C}$ and let $S_j (\alpha_j )$ be the
Gaussian state (\ref{3-1}) with
\begin{equation}\label{4-1}     \mbox{Tr}S_j (\alpha_j )a_j =\alpha_j ,\qquad
\mbox{Tr}S_j (\alpha_j )a_j^{\dagger}a_j = N_j + |\alpha_j |^2 . \end{equation}
Denoting $\alpha = (\alpha_j)$ we consider the tensor product states
\begin{equation}\label{4-2}
S_\alpha = \otimes_j S_j (\alpha_j ) ,
\end{equation}
and we are interested in the memoryless quantum Gaussian channel $\alpha
\rightarrow S_{\alpha}$ satisfying
the additive constraint (\ref{2-1}), where $f$ is the energy of the signal
$$f(\alpha ) =\sum_j \hbar \omega_j |\alpha_j |^2 .$$ Note that according to
(\ref{3-4}), the entropy of the states (\ref{4-2}) is equal to
\begin{equation}\label{4-3}
H ( S_\alpha ) = H (S_0) = \sum_j H (S_j(0)) = \sum_j g (N_j).\end{equation}
We shall denote by
\begin{equation}\label{Planck}N_j (\theta ) =  {1 \over \mbox{e}^{\theta \hbar\omega_j} -1}\end{equation}
the Planck distribution which maximizes the entropy (\ref{4-3}) under the
constraint $$\sum_j \hbar\omega_i N_j \leq \mbox{const},$$ and remark that
\begin{equation}\label{4-14}    g(N_j (\theta )) =  g\left( {1 \over \mbox{e}^{\theta \hbar\omega_j}-1}\right) = {\theta\hbar\omega_j \over
\mbox{e}^{\theta\hbar\omega_j} - 1} - \log ( 1-
\mbox{e}^{-\theta\hbar\omega_j} ).\end{equation}

Let ${\cal P}_1$ be the set of probability distributions $\pi (d^2
\alpha )$ satisfying
\begin{equation}\label{4-4}
\int\ \left(\sum_j\hbar \omega_j |\alpha_j |^2 \right)\pi (d^2\alpha ) \leq  E .
\end{equation}
We also use the notation $(y)_+ =
\max (y, 0)$.

{\bf Proposition 3}. {\sl The capacity of the memoryless quantum Gaussian
channel with the additive constraint onto the signal energy is equal to
\begin{equation}\label{4-5}
C = \sum_{j}\left(g(N_j(\theta )) - g(N_j)\right )_+ ,\end{equation} where
$\theta$ is chosen such that} \begin{equation}\label{4-6}
\sum_{j}\hbar\omega_j (N_j (\theta ) - N_j)_+ = E  .\end{equation}

{\sl Proof}. Proposition 2 applies, so we have only to calculate supremum in
the right hand side of (\ref{2-6}) with ${\cal P}_1$ given by (\ref{4-4}).
Let us show that it is achieved on the Gaussian probability distribution
\begin{equation}\label{4-7}
\pi (d^2 \alpha) = \mbox{exp}\left(-\sum_j{|\alpha_j|^2 \over m_j^*}\right)
\prod_j d^2 \alpha_j ,\end{equation}
where
\begin{equation}\label{4-8}  m_j^* = \left(N_j (\theta ) -N_j\right)_+ ,   \end{equation}
and $\theta$ is chosen such that
\begin{equation}\label{4-9}\sum_j\hbar\omega_j m_j^* = E.
\end{equation}
(If $m_j^{*}=0$, we have in mind in (\ref{4-7}) the Gaussian distribution
degenerated at $0$.)

Since $\Delta H (\pi) = H({\bar S}_{\pi}) - H(S_0)$, we have by (\ref{4-3})
and subadditivity of quantum entropy
\begin{equation}\label{4-11}
\Delta H (\pi) \leq\sum_j\Delta H (\pi^{(j)}),\end{equation}
where $\pi^{(j)}$ are the marginal distributions of $\pi$.

Let us denote $|\alpha_j |^2 = m_j$. We first maximize with respect to
$\pi^{(j)}$ satisfying $$\int\hbar \omega_j |\alpha_j |^2 \pi^{(j)}
(d^2\alpha_j ) = m_j ,$$ and then with respect to $m_j$ satisfying
\begin{equation}\label{4-12}m_j\geq 0 ,\qquad
\sum_j\hbar\omega_j m_j \leq E .
\end{equation}
According to \S 1, the first maximization is achieved by the Gaussian
probability distribution $$\pi^{(j)} (d^2 \alpha_j) =
\mbox{exp}\left(-{|\alpha_j|^2 \over m_j}\right)
d^2 \alpha_j $$ We can then take $$\pi (d^2 \alpha) = \prod_j
\pi^{(j)} (d^2 \alpha_j), $$ for which equality holds in the second relation of (\ref{4-12}).
For such $\pi $
\begin{equation}\label{4-13}
\Delta H (\pi) = \sum_j \left[g(N_j+m_j) - g(N_j)\right].
\end{equation}

We have thus arrived at maximizing (\ref{4-13}) under the constraint
(\ref{4-12}), which is similar to the problem of finding the capacity of
quasiclassical multimode photon channel, considered in \cite{xle}.  The
Kuhn-Tucker conditions for the solution $m_j^*$ of this problem imply
equations (\ref{4-8}), (\ref{4-9}), and the last one is the same as
(\ref{4-6}).  $\Box$

In the case of oscillators in thermal equilibrium \cite{xle},
\cite{xbowen},
\cite{xcaves}, $N_j$ themselves are given by the
Planck distribution $$N_j = N_j (\theta_P )\equiv {1 \over \mbox{e}^{\theta_P
\hbar\omega_j} -1},$$
where $\theta_P$ is such that $$\sum_{j}\hbar\omega_j N_j =
\sum_{j}{\hbar\omega_j \over
\mbox{e}^{\theta_P\hbar\omega_j} - 1} = P$$ is the mean energy of the
oscillator noise. The entropy (\ref{4-3}) of the signal states is $$ s(P) =
\sum_j g(N_j(\theta_P)) = \sum_{j}\left\{{\theta_P\hbar\omega_j \over
\mbox{e}^{\theta_P\hbar\omega_j} - 1} - \log ( 1-
\mbox{e}^{-\theta_P\hbar\omega_j  })\right\}.$$
In this case the formula (\ref{4-5}) becomes
\begin{equation}\label{4-15}C = s(P+E) - s(P).
\end{equation}
In particular for the pure state Gaussian channel, where $N_j\equiv 0, P=0$
one has
\begin{equation}\label{4-16} C = s(E).
\end{equation}

\vskip20pt
\centerline{\bf \S 4. The quantum Gaussian waveform channel}
\vskip10pt

We now pass to consideration of more realistic dynamical model of the
Gaussian channel -- that of the waveform channel. In classical information
theory the waveform channel is treated by reduction to parallel channels, i.
e. by decomposing the Gaussian stochastic process into independent
one-dimensional modes. In quantum theory such a decomposition plays an
additional role as a tool for {\sl quantization} of the classical stochastic
process. The partly heuristic procedure described below is an analog of
classical decomposition into harmonic modes (\cite{xgal}, \S 8.3).  A
mathematical formulation in terms of a {\sl quantum stochastic process}
avoiding this procedure is possible (see the end of this Section), but
rigorous calculation of the capacity based on this formulation remains an
open problem.

Let us consider the periodic operator-valued function
\begin{equation}\label{5-1}
X(t) = \sum_j\sqrt{2\pi\hbar\omega_j \over T}\left(a_j\mbox{e}^{-i\omega_j t}
+a_j^{\dagger}\mbox{e}^{i\omega_j t}\right) ,\quad t\in [0,T],
\end{equation}
where $[0,T]$ is the observation interval, \begin{equation}\label{5-2}
\omega_j = {2\pi j \over T}, \qquad
j= 1,2,...\end{equation} are the oscillator frequencies and $a_j^{\dagger},
a_j$ are the creation-annihilation operators. In quantum electrodynamics a
similar relation represents a component of the electric field in a planar
wave with periodic boundary conditions on finite interval (see e.g.
\cite{xhel}).
To avoid problems related to infinite degrees of freedom, we shall restrict
summation in (\ref{5-1}) to a finite range $I_T$ which will grow with $T$. In
the band-limited case, where $0< {\underline \omega}<\omega<{\bar \omega}
<\infty$, we can put $$I_T = \{j: {\underline \omega}\leq\omega_j\leq{\bar
\omega}\}.$$
In the case of infinite frequency band (where $ {\underline \omega}=0,\quad
{\bar \omega}=\infty$), we shall take $$I_T = \{j: {\underline
\omega}_T\leq\omega_j\leq{\bar \omega}_T\},$$
where ${\underline \omega}_T\downarrow 0 ,\quad {\bar
\omega}_T\uparrow\infty.$

We have
\begin{equation}\label{5-3}
{1 \over 4\pi} \int_0^T X(t)^2 dt = \ \sum_j \hbar\omega_j\left(
a_j^{\dagger}a_j +{1 \over 2}\right)
\end{equation}
for the corresponding energy operator.

We assume that the mode $a_j$ is described by the Gaussian state $S_j
(\alpha_j)$ with the first two moments given by (\ref{4-1}), so that the
whole process $X(t)$ is characterized by the product Gaussian state
(\ref{4-2}), such that $$
\mbox{Tr} S_\alpha X(t) = \alpha (t) ,$$\begin{equation}\label{5-4}
\mbox{Tr}S_\alpha
{1 \over 4\pi} \int_0^T X(t)^2 dt = \sum_j \hbar\omega_j\left( N_j +{1 \over
2}\right)+{1 \over 4\pi} \int_0^T \alpha(t)^2 dt,
\end{equation}
where
\begin{equation}\label{5-5}
\alpha (t)=\sum_j\sqrt{2\pi\hbar\omega_j \over T}\left(
\alpha_j\mbox{e}^{-i\omega_j t}
+{\bar \alpha_j}\mbox{e}^{i\omega_j t}\right),
\end{equation}
\begin{equation}\label{5-6}
{1 \over 4\pi} \int_0^T \alpha(t)^2 dt = \sum_j \hbar\omega_j |\alpha_j|^2 .
\end{equation}
The signal is thus represented by the real function $\alpha(t)$ and the mean
power constraint on the signal is as follows
\begin{equation}\label{5-7}
{1 \over 4\pi} \int_0^T \alpha(t)^2 dt \leq T E .
\end{equation}

A code $(W,X)$ for such a channel is a collection $(\alpha^{1}(\cdot), X_1 ),
...,(\alpha^{M}(\cdot), X_M)$, where $\alpha^{k}(\cdot)$ are functions
of $t\in [0,T]$ representing different signals; one defines the capacity of
the channel as supremum of values of $R$ for which the infimum of the average
error probability
\begin{equation}\label{5-8}
{\bar \lambda}(W,X) = {1 \over M}\sum_{j=1}^{M}(1-\mbox{Tr}S_{\alpha^{j}}
X_j).
\end{equation}
with respect to all codes of the size $M = \mbox{e}^{T R}$ tends to zero as
$T\rightarrow\infty$.

{\bf Proposition 4}. {\sl Assume that $N_j=N(\omega_j)$, where $N(\omega)$ is
a continuous function. The capacity of the Gaussian waveform channel as
defined above exists and is equal to
\begin{equation}\label{5-9}
C = {1 \over 2\pi }\int_{\underline \omega}^{\bar \omega}(g(N_{\theta}
(\omega )) - g(N(\omega )))_+ d\omega , \end{equation} where
\begin{equation}\label{5-10}
N_{\theta}(\omega )= {1 \over
\mbox{e}^{\theta\hbar\omega} - 1} ,     \end{equation}
and $\theta$ is chosen such that}
\begin{equation}\label{5-11}
{1 \over 2\pi }\int_{\underline \omega}^{\bar \omega}\hbar\omega
(N_{\theta}(\omega ) - N(\omega ))_+ d \omega = E .  \end{equation}

{\sl Proof}.  We start with considering the band-limited case and first prove
that $\inf_{W,X}{\bar \lambda} (W,
\\ X)\not\rightarrow 0$ for $R>C$.
From the Fano inequality of the type (\ref{1a-10}) and Proposition 3 we have
\begin{equation}\label{5-12}
TR\cdot (1 -\inf_{W,X}{\bar \lambda}(W, X))\leq C_T +1 , \end{equation} where
\begin{equation}\label{5-13}     C_T = \max\sum_{j\in I_T}[g(N_j +m_j) - g(N_j )], \end{equation}
and the maximum is taken over the set $$m_j\geq 0, \qquad {1 \over 2\pi
}\sum_{j\in I_T}\hbar\omega_j m_j \Delta\omega_j\leq E ,$$ with
\begin{equation}\label{5-14}
\Delta\omega_j ={2\pi \over T} .
\end{equation} Introducing the piecewise constant function
$$N_T(\omega)=N_j,\quad \omega_{j-1}< \omega< \omega_j,$$ we have $${C_T
\over T}\leq{1 \over 2\pi}\max_{{\cal M}_({\underline \omega},{\bar \omega})}\int_{\underline \omega}^{\bar \omega}
[g(N_T ( \omega)+m(\omega)) - g(N_T(\omega) )]d\omega,$$ where $${\cal M}
({\underline \omega},{\bar \omega})=\{m(\cdot):m(\omega)\geq 0;\quad {1 \over
2\pi }\int_{\underline \omega}^{\bar \omega}
\hbar \omega m( \omega)d\omega\leq E \}.$$
Since $N( \omega)$ is continuous, it is uniformly continuous on $[{\underline
\omega}, {\bar \omega}]$ and therefore $N_T( \omega)$
tends uniformly to $N( \omega)$ as $T\rightarrow\infty$. It follows that $${
\limsup}_{T\rightarrow\infty}{C_T \over T}\leq\max_{{\cal M}_({\underline \omega},{\bar \omega})}{1 \over 2\pi}
\int_{\underline \omega}^{\bar \omega}
[g(N ( \omega)+m(\omega)) - g(N(\omega) )]d\omega.$$ However, this maximum is
achieved on
\begin{equation}\label{5-15}
m^{*} (\omega) = \left(N_{\theta}(\omega ) - N(\omega)\right)_+ ,
\end{equation}
and is equal to $C$ as defined by (\ref{5-9})-(\ref{5-11}). Therefore from
(\ref{5-12}) we conclude that $\inf_{W,X}{\bar \lambda} (W,X)\not\rightarrow
0$ for $R>C$.

We now show that the average error probability tends to zero if $R<C$.  Let
$\pi (d^2 \alpha )$ be the Gaussian probability distribution (\ref{4-7}),
where $m_j^{*}$ are given by (\ref{4-8}) with $\theta =\theta_T$ chosen such
that
\begin{equation}\label{5-16}
{1 \over 2\pi }\sum_{j\in I_T}\hbar\omega_j m_j^{*}\Delta\omega_j = E .
\end{equation}
Applying the basic bound (\ref{gs10}) with the word length $n=1$ and with
$\delta$ replaced by $\delta T$, we have
\begin{equation}\label{5-17}
\inf_X {\bar \lambda}(W,X)\leq\end{equation}  $$
\leq {1 \over M}\sum_{j=1}^{M}\left\{4\mbox{Tr}S_{\alpha^{j}}(I-P)
+ 4\mbox{Tr}S_{\alpha^{j}} (I-P_{\alpha^{j}})+ \sum_{k\neq
j}\mbox{Tr}PS_{\alpha^{j}}PP_{\alpha^{k}}\right\}, $$ where $P$ is the
spectral projection of ${\bar S}_{\pi}$ corresponding to the eigenvalues in
the range $(\mbox{e}^{-[H({\bar S}_{\pi})+\delta T]},\\ \mbox{e}^{-[H({\bar
S}_{\pi})-\delta T]})$, and $P_{\alpha }$ is the spectral projection of $
S_{\alpha}$ corresponding to the eigenvalues in the range $(\mbox{e}^{-[{\bar
H}( S_{(\cdot)})+\delta T]},
\mbox{e}^{-[{\bar H}( S_{(\cdot)})-\delta T]})$.
Since $S_\alpha $ are unitary equivalent to $S_0$, then ${\bar H}(
S_{(\cdot)}) = H(S_0)$ and the last term in (\ref{5-17}) is simply
\begin{equation}\label{5-18}
\mbox{Tr}S_0(I-P_0),
\end{equation}
which is similar to \begin{equation}\label{5-19} \mbox{Tr}{\bar
S}_{\pi}(I-P). \end{equation}

We wish to estimate the terms (\ref{5-18}), (\ref{5-19}) for the Gaussian
density operators $S_0, {\bar S}_{\pi}$.  For definiteness let us take
(\ref{5-18}). We have
\begin{equation}\label{5-20}
\mbox{Tr}S_0(I-P_0) = {\sf Pr}\left\{ |- \log\lambda_{(\cdot)} - H(S_0)| \geq\delta T
\right\},
\end{equation} where ${\sf Pr}$ is the distribution of eigenvalues
$\lambda_{(\cdot)}$ of $S_0$. By Chebyshev inequality, this is less or equal
to ${\sf D}(\log\lambda_{(\cdot)} )/ \delta^{2} T^{2}$. Now ${\sf
D}(\log\lambda_{(\cdot)} )=\sum_j{\sf D}_j(\log\lambda_{(\cdot)} )$, where
${\sf D}_j$ is the variance of $\log\lambda_{(\cdot)}$ for the $j$-th mode.
From (\ref{3-3}) we see that the eigenvalues of $S_j (0)$ are $$\lambda_n^j =
{N_j^{n} \over (N_j+1)^{n+1}}; \qquad n=0,1,...,$$ hence
\begin{equation}\label{5-21}
{\sf D}_j(\log\lambda_{(\cdot)} )=\sum_{n=0}^{\infty}( -\log\lambda_n^j -
H(S_0))^{2}
\lambda_n^j      \end{equation}\begin{equation}
=\log^{2}{N_j +1 \over N_j} \sum_{n=0}^{\infty}(n-N_j)^{2} {N_j^{n} \over
(N_j+1)^{n+1}} =F(N_j), \end{equation} where $$F(x) =x (x+1)\log^{2}{x +1
\over x} $$
is a bounded function on $(0, \infty)$.  Thus finally
\begin{equation}\label{5-23}
\mbox{Tr}S_0(I-P_0)\leq {\sum_j F(N_j) \over
\delta^{2}T^{2}},
\end{equation} and a similar estimate holds for $ \mbox{Tr}{\bar
S}_{\pi}(I-P)$ with $N_j$ replaced with $N_j + m_j^*$.

Now let the words $\alpha^{1},...,\alpha^{M}$ be taken randomly with the
joint probability distribution $\tilde {\sf P}$ defined similarly to
(\ref{1a-11}) starting from the probability distribution $ {\sf P}$ with
respect to which the words are independent and have the same probability
distribution $\pi (d^{2}\alpha )$.  Then ${\tilde {\sf E}} \xi \leq 2^m {\sf
E} \xi$ for any nonnegative random variable $\xi$ depending on $m$ words.
Therefore from (\ref{5-17}) $${\tilde {\sf E}}\inf_X {\bar \lambda}(W,X)\leq
{1 \over M}\sum_{j=1}^{M}\left\{8{\sf
E}\mbox{Tr}S_{\alpha^{(j)}}(I-P)+4\mbox{Tr}S_0 (I-P_0) + \sum_{k\neq j}4 {\sf
E}\mbox{Tr}PS_{\alpha^{(j)}}P P_{\alpha^{(k)}}\right\}$$ $$= 8\mbox{Tr}{\bar
S}_{\pi}(I-P) + 4\mbox{Tr}S_0 (I-P_0)+4 (M-1)\mbox{e}^{-[\Delta H(\pi)-2\delta
T]}$$ $$\leq{8\sum_{j\in I_T}F(N_j+m_j^{*} )
\over \delta^{2}T^{2}} +{4\sum_{j\in I_T}F(N_j)
\over \delta^{2}T^{2}} +4 (M-1)\mbox{e}^{-[C_T -2\delta T]}.$$
Since the function $F(x)$ is bounded and the size of $I_T$ is proportional to
$T$, the sums in the first two terms have the order $T$, the terms themselves
having the order $T^{-1}$. To complete the proof we have only to show that
\begin{equation}\label{5-24}
\liminf_{T\rightarrow\infty}{C_T \over T}\geq C.
\end{equation}

Let $m^{*}(\omega)$ be the function (\ref{5-15}), and let $\omega_j'$ be the
point on the segment $[\omega_{j-1},\omega_j]$ at which it achieves its
minimum, then $${1 \over 2\pi}\sum_{j\in I_T}\hbar \omega_j' m^{*}(\omega_j')
\leq{1 \over 2\pi}\int_{\underline \omega}^{\bar \omega}
\hbar \omega m^{*}(\omega)d\omega = E,$$
hence $${C_T \over T}\geq{1 \over 2\pi}\sum_{j\in I_T}[g(N(\omega_j)
+m^*(\omega_j')) - g(N(\omega_j) )]\Delta\omega_j .$$ Since both $N(\omega) $
and $m^{*}(\omega)$ are continuous, the last sum tends to $$\int_{\underline
\omega}^{\bar \omega}[g(N(\omega) +m^*(\omega)) -
g(N(\omega) )]d\omega =C,$$ and the proof is completed.

We now turn to the case of the infinite frequency band $(0,\infty)$. By
applying argument similar to given above, one sees it is sufficient to show
that
\begin{equation}\label{5-25}
\lim_{T \to \infty}{C_T \over T} = C(0,\infty)\equiv\max_{{\cal M}(0,\infty)}{1 \over 2\pi}
\int_{0}^{\infty}[g(N(\omega) +m(\omega)) -
g(N(\omega) )]d\omega ,
\end{equation}
where
\begin{equation}\label{5-26}
{\cal M}(0,\infty) = \{m(\cdot): m(\omega)\geq 0,{1 \over
2\pi}\int_{0}^{\infty}\hbar\omega m(\omega) d\omega \leq E\}. \end{equation}
The maximum is achieved on the function $m^{*}(\omega)$ of the form
(\ref{5-15}), where $\theta$ is such that $${1 \over
2\pi}\int_{0}^{\infty}\hbar\omega m^{*}(\omega) d\omega = E.$$

Let us take $0<{\underline \omega}<{\bar \omega}<\infty$. By omitting
frequencies outside this band, we obtain $$\liminf_{T \to \infty}{C_T \over
T} \geq\max_{{\cal M}({\underline \omega},{\bar \omega})}{1 \over
2\pi}\int_{{\underline \omega}}^{\bar \omega}[g(N(\omega) +m(\omega)) -
g(N(\omega) )]d\omega $$ $$\geq{1 \over 2\pi}\int_{{\underline \omega}}^{\bar
\omega}[g(N(\omega) +m^{*}(\omega)) - g(N(\omega) )]d\omega ,
$$ since $m^{*}(\cdot )\in{\cal M}({\underline \omega},{\bar \omega})$.
Taking the limit as ${\underline \omega}\rightarrow 0,{\bar
\omega}\rightarrow\infty$,
we prove the $\geq$ part of (\ref{5-25}).

To prove the $\leq$ part, we consider the relation
\begin{equation}\label{5-27}
{C_T \over T}={1 \over 2\pi}\sum_{j}[g(N_j + m_j^{*})-g(N_j)]\Delta\omega_j,
\end{equation}
where $$ m_j^{*} =\left({1 \over
\mbox{e}^{\theta_T\hbar\omega_j}-1}-N_j\right)_+$$
and $\theta_T$ is chosen in such a way that
\begin{equation}\label{5-28}    {1 \over 2\pi} \sum_{j}\hbar\omega_j m_j^{*}\Delta\omega_j =E. \end{equation}
By considering the piecewise constant functions $$N_T (\omega)=N_j ,\qquad
m_T (\omega)=m_j^{*}\qquad \mbox{for} \quad\omega_{j-1}<\omega
\leq\omega_j$$
we can write the right hand side of (\ref{5-27}) as $${1 \over
2\pi}\int_{0}^{\infty}[g(N_T(\omega) +m_T(\omega)) - g(N_T(\omega) )]d\omega
$$ $$= {1 \over 2\pi}\int_{0}^{\infty}[g(N(\omega) +m_T(\omega)) -
g(N(\omega) )]d\omega$$ $$+{1 \over 2\pi}\int_{0}^{\infty}[g(N_T(\omega)
+m_T(\omega)) - g(N(\omega) +m_T(\omega)) + g(N(\omega)-g(N_T(\omega)
)]d\omega. $$ Taking into account that $$ {1 \over
2\pi}\int_{0}^{\infty}\hbar\omega m_T(\omega) d\omega \leq {1 \over
2\pi}\sum_{j}\hbar\omega_j m_j^{*}
\Delta\omega_j =E,$$
we see that the first term is less or equal to $$\max_{{\cal M}(0,\infty)}{1
\over 2\pi}
\int_{0}^{\infty}[g(N(\omega) +m(\omega)) -
g(N(\omega) )]d\omega = C(0,\infty).$$ It remains to show that the second
term tends to zero. We shall show it by using the Lebesgue dominated
convergence theorem.  Since $N(\omega)$ is continuous,
$N_T(\omega)\rightarrow N(\omega)$ and $ g(N_T(\omega))\rightarrow
g(N(\omega))$ pointwise. Next we observe that $\theta_T$ is separated from
$0$ as $T\rightarrow \infty$, that is $\theta_T\geq\theta_0>0$. Indeed,
assume that $\theta_T\downarrow 0$ for some sequence $T\rightarrow \infty$,
then the sequence of continuous functions $$\left({1 \over
\mbox{e}^{\theta_T\hbar\omega}-1}-N(\omega)\right)_+$$
converges to $\infty$ uniformly in every interval $0<{\underline
\omega}\leq\omega\leq{\bar \omega}<\infty$, which contradicts to the condition
(\ref{5-28}). It follows that for any $\omega>0$ the quantity $$N(\omega)
+m_T(\omega) =\max\left({1 \over \mbox{e}^{\theta_T\hbar\omega} -1}, N
(\omega)\right)$$ is bounded as $T\rightarrow \infty$. Since $g(\omega)$ is
uniformly continuous on any bounded interval, it follows that $$g(N_T(\omega)
+m_T(\omega)) - g(N(\omega) +m_T(\omega))\rightarrow 0.$$

It remains to show that the integrand is dominated by an integrable function.
Taking into account that $h''(x)\leq 0$ for $x\geq 0$, we deduce that
$g(x+y)-g(x)\leq g(y)$ for $x,y\geq 0$. Therefore the integrand is dominated
by the function $2g(m_T(\omega ))$. But $$m_T(\omega )\leq {1 \over
\mbox{e}^{\theta_T\hbar\omega}-1}\leq{1 \over \mbox{e}^{\theta_0\hbar\omega}-1}. $$ Thus
$$g(m_T(\omega ))\leq g\left({1 \over
\mbox{e}^{\theta_0\hbar\omega}-1}\right)=
{\theta_0\hbar\omega \over \mbox{e}^{\theta_0\hbar\omega}-1}-\log (
1-\mbox{e}^{-\theta_0\hbar\omega}),$$ which is positive integrable
function.$\Box$

These formulas take especially nice form for the equilibrium noise when
$N(\omega) = N_{\theta_P}(\omega )\equiv (\mbox{e}^{\theta_P\hbar\omega} -
1)^{-1}$ with $\theta_P$ determined from $$ {1 \over 2\pi
}\int_0^{\infty}{\hbar\omega\over
\mbox{e}^{\theta_P\hbar\omega} - 1} d \omega = P.$$ By using the formula
$$\int_0^{\infty} {x\over \mbox{e}^{x}-1}dx = {\pi^{2} \over 6},$$ one finds
$\theta_P = \sqrt{\pi / 12\hbar P}$, and $$ s(P) = {1 \over 2\pi
}\int_0^{\infty}g(N_{\theta_P}(\omega ))d\omega $$
\begin{equation}\label{5-29} = {1 \over 2\pi }\int_0^{\infty}\left\{{\theta_P\hbar\omega \over
\mbox{e}^{\theta_P\hbar\omega} - 1} - \log ( 1-
\mbox{e}^{-\theta_P\hbar\omega}) \right\} d\omega = {\pi \over 6\theta_P \hbar}
= \sqrt{\pi P \over 3\hbar}, \end{equation} whence
\begin{equation}\label{5-40}
C(0,\infty)=\sqrt{\pi (P+E) \over 3\hbar}-\sqrt{\pi P \over 3\hbar},
\end{equation}
which coincides with the capacity of the infinite band photon channel
calculated in
\cite{xle}. For the pure state channel
\begin{equation}\label{5-41}
C(0,\infty)=\sqrt{\pi E \over 3\hbar} , \end{equation} a formula found also
in \cite{xbowen}.

Let us now formulate the problem more in the spirit of the treatment of
Gaussian noise in classical information theory. In the limit $T\rightarrow
\infty$ one expects that the
periodic process (\ref{5-1}) turns into the ``signal $+$ noise'' process
$$X(t) = \alpha (t) + Y(t),\qquad t\geq 0,$$ where $\alpha$ is the classical
signal and $Y(t)$ is the quantum Gaussian noise
\begin{equation}\label{5-42}
Y(t) = \int_{0}^{\infty}\sqrt{\hbar\omega}\left(
dA_{\omega}\mbox{e}^{-i\omega t} + dA_{\omega}^{\dagger}\mbox{e}^{i\omega
t}\right).  \end{equation} Here $A_{\omega}$ is the quantum Gaussian
independent increment process having the commutator $$[dA_{\omega},
dA_{\lambda}^{\dagger} ] = \delta (\omega - \lambda ) d\omega d\lambda ,$$
zero mean, and the correlation $$\langle dA_{\omega}^{\dagger} dA_{\lambda}
\rangle= \delta (\omega -  \lambda )  N(\omega )d\omega d\lambda$$ (all other commutators and correlations vanish).

By using this with (\ref{5-42}) one obtains the noise commutator
\begin{equation}\label{5-43}    [Y(t) , Y(s)] = 2i\hbar \int_0^{\infty}\omega\sin\omega (s-t) d\omega =
2 i\hbar\pi\delta ' (t-s) , \end{equation} and the noise correlation function
\begin{equation}\label{5-44}     \langle Y(t) Y(s) \rangle = B (t-s) + K (t-s) , \end{equation}  with
$$B (t) = 2\hbar\int_0^{\infty}\omega N(\omega )\cos\omega t d\omega $$ and
$$ K (t) = \hbar \int_0^{\infty}\omega\mbox{e}^{- i\omega t}d\omega = - \hbar
[t^{-2} - i\pi\delta ' (t)]$$ is the zero temperature correlation.

The process $X(t)$ is observed on the time interval $[0, T]$ which means that
one considers the Gaussian (quasifree) state with mean $\alpha (t)$ and the
correlation function (\ref{5-44}) on the algebra of canonical commutation
relation generated by $X(t); t\in [0, T]$, which is determined by the
commutator (\ref{5-43}) (see e.g. \cite{xstat}).  One imposes the power
constraint (\ref{5-7}) and defines the capacity in the same way as we have
done before Proposition 4. The proof makes it plausible that the capacity of
the ``signal $+$ noise'' process is just one given by that Proposition.
However an attempt to prove this following the classical method of reduction
to parallel channels \cite{xgal} meets the following new difficulties.  A
minor problem is that the kernels (\ref{5-43}), (\ref{5-44}) are now
generalized functions.  More important is that in the classical case one has
two quadratic forms: correlation and energy (which is just the $L^2$ inner
product), that are simultaneously diagonalized by solving the integral
equation with the kernel (\ref{5-44}). In quantum case one has additional
skew-symmetric form -- the commutator -- which also should be transformed to
a canonical representation allowing decomposition into parallel channels.
However this is not always possible. The proof of Proposition 4 shows that in
a sense this happens asymptotically (as $ T\rightarrow\infty$), but a
rigorous proof of that is still lacking.
\vskip20pt
{\sc VI. Some open problems}
\vskip10pt
Several questions remain, some of which were mentioned in the text.  Let us
remind them adding few further problems and comments:

1) Superadditivity of the entropy bound for general quantum channel
\cite{xbennet}; our
conjecture is that if channels with this property at all exist, they might
be found in a neighbourhood of the identity channel. The perturbation of the
identity should be truly quantum and irreversible and small enough to enable
neglecting probability of more than one error in the product channel. One
then may try to find input states exhibiting the superadditivity property by
using quantum codes correcting one error
\cite{xcal}, \cite{xste};

2) Finding practical block codes with substantial gain from the strict
superadditivity of the capacity \cite{xsasa};

3) Exponential upper bound for error probability in c-q channel for general
signal states, allowing for lower bound of the quantum reliability function,
see \cite{xbur};

4) Lower bound for error probability at least for pure state channel (an
analog of sphere-packing bound), see \cite{xbur};

5) Consistent treatment of quantum Gaussian waveform channel as described at
the end of \S V.4.

All these problems address transmission of classical information through
quantum channels. There is yet ``more quantum'' domain of problems concerning
reliable transmission of entire quantum states under a given fidelity
criterion \cite{xben}.  The very definition of the relevant ``quantum
information'' is far from obvious. Important steps in this direction were
made in \cite{xbarnum} , where in particular a tentative converse of the
relevant coding theorem was suggested.  However the proof of the
corresponding direct theorem remains an open question.

{\sc Acknowledgements}. The author acknowledges hospitality of Prof. O. Hirota 
(Research Center of
Tamagawa University), Prof.
O. Melsheimer (University of Marburg) and Prof. L. Lanz (University of
Milan), where parts of this work were written. The work was partially 
supported by RFBR, JSPS, DAAD and  Italian Ministry for Foreign Affairs
under a convention with Landau Network - Centro Volta.

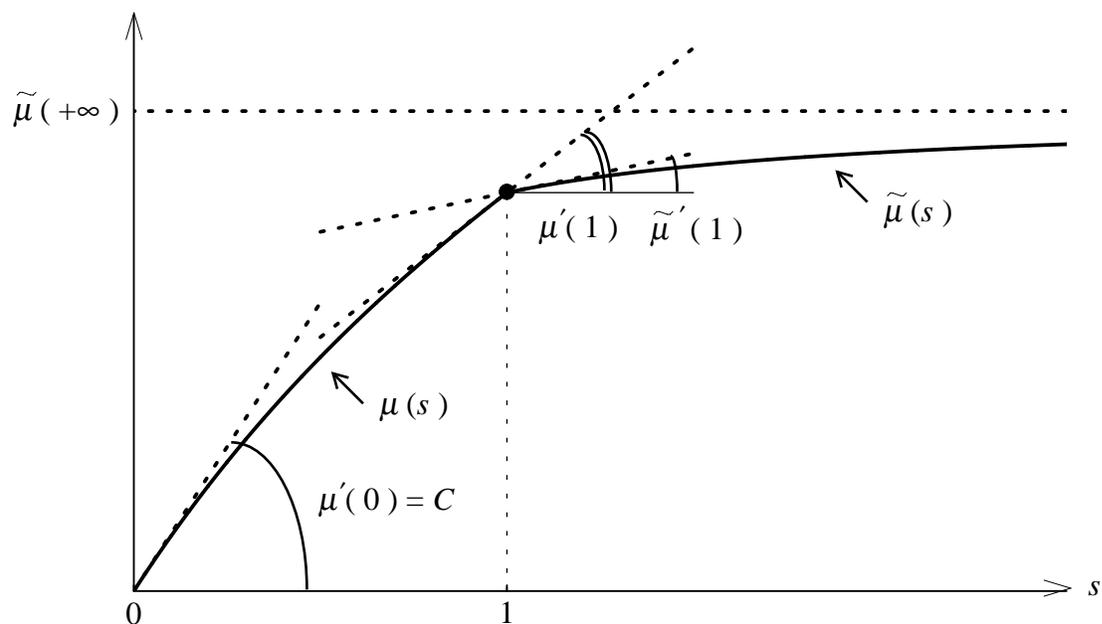
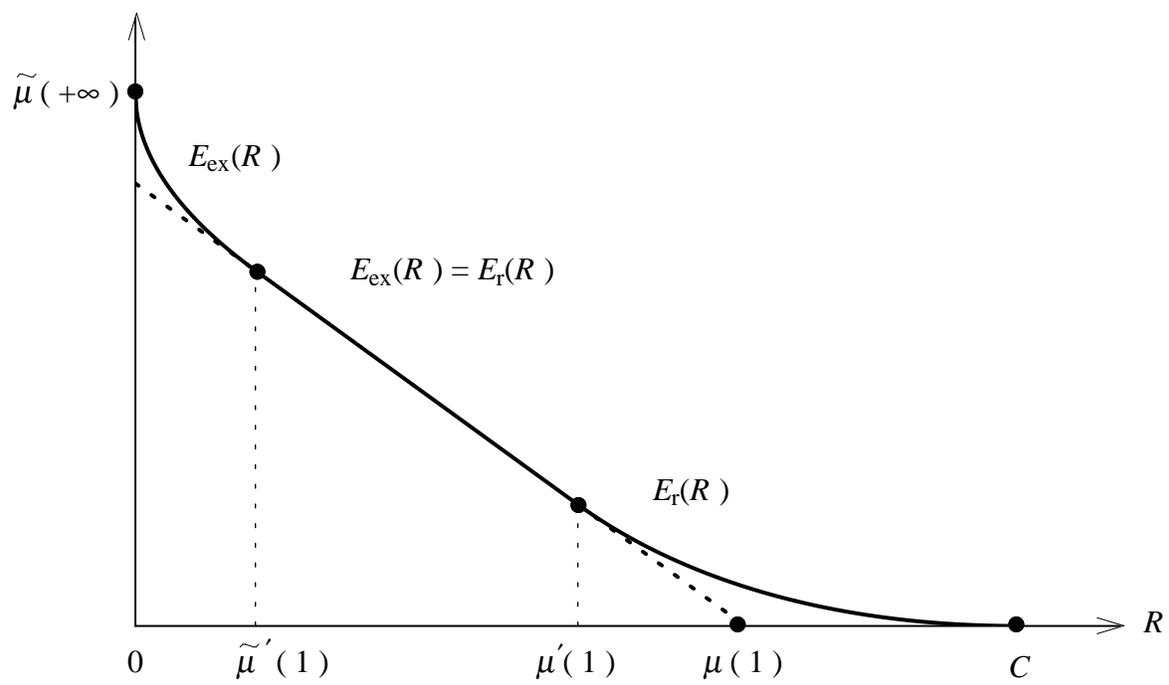
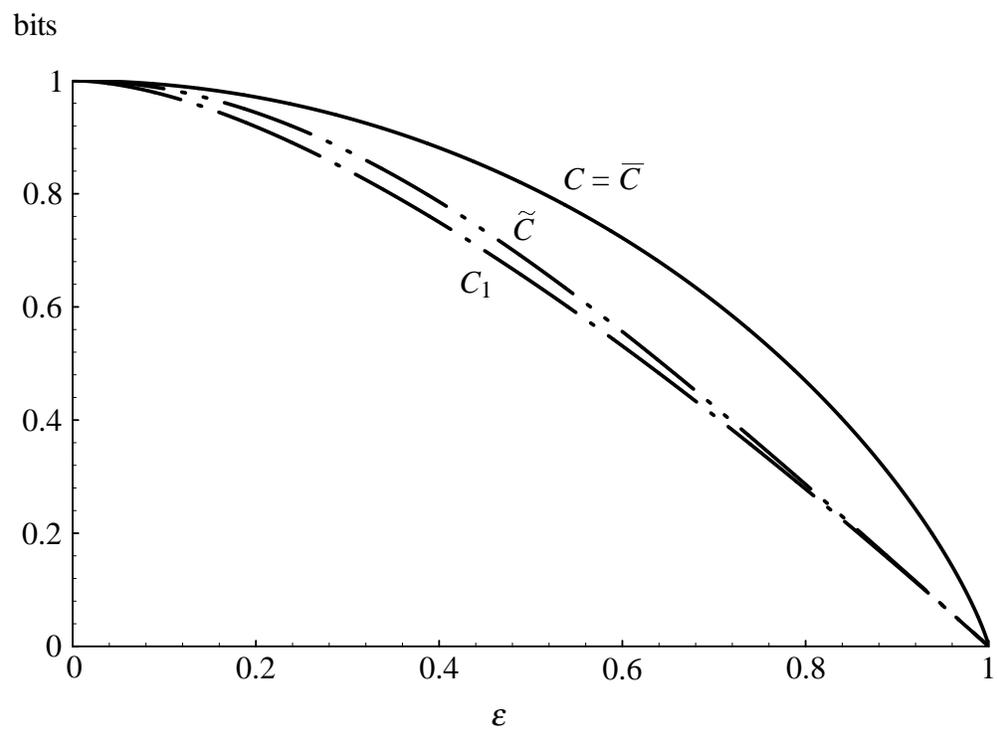
\end{document}